\documentclass[12pt]{article}
\catcode`\@=11
\def\@normalsize{\@setsize\normalsize{15pt}\xiipt\@xiipt
\abovedisplayskip 14pt plus3pt minus3pt%
\belowdisplayskip \abovedisplayskip
\abovedisplayshortskip  \z@ plus3pt%
\belowdisplayshortskip  7pt plus3.5pt minus0pt}
\def\small{\@setsize\small{13.6pt}\xipt\@xipt
\abovedisplayskip 13pt plus3pt minus3pt%
\belowdisplayskip \abovedisplayskip
\abovedisplayshortskip  \z@ plus3pt%
\belowdisplayshortskip  7pt plus3.5pt minus0pt
\def\@listi{\parsep 4.5pt plus 2pt minus 1pt
            \itemsep \parsep
            \topsep 9pt plus 3pt minus 3pt}}
\def\underline#1{\relax\ifmmode\@@underline#1\else
        $\@@underline{\hbox{#1}}$\relax\fi}
\@twosidetrue
\relax
\catcode`@=12
\catcode`\@=11

\def\section{\@startsection{section}{1}{\z@}{3.5ex plus 1ex minus
   .2ex}{2.3ex plus .2ex}{\large\bf}}

\def\ps@headings{\def\@oddfoot{}\def\@evenfoot{}
\def\@oddhead{\hbox{}\hfill
        \makebox[.5\textwidth]{\raggedright\ignorespaces --\thepage{}--
        \hfill }}
\def\@evenhead{\@oddhead}
\def\subsectionmark##1{\markboth{##1}{}}
}
\ps@headings
\catcode`\@=12
\relax

\def\figcap{\section*{Figure Captions\markboth
        {FIGURECAPTIONS}{FIGURECAPTIONS}}\list
        {Fig. \arabic{enumi}:\hfill}{\settowidth\labelwidth{Fig. 999:}
        \leftmargin\labelwidth
        \advance\leftmargin\labelsep\usecounter{enumi}}}
 \relax
\def\tablecap{\section*{Table Captions\markboth
        {TABLECAPTIONS}{TABLECAPTIONS}}\list
        {Table \arabic{enumi}:\hfill}{\settowidth\labelwidth{Table 999:}
        \leftmargin\labelwidth
        \advance\leftmargin\labelsep\usecounter{enumi}}}
 \relax
\def\reflist{\section*{References\markboth
        {REFLIST}{REFLIST}}\list
        {[\arabic{enumi}]\hfill}{\settowidth\labelwidth{[999]}
        \leftmargin\labelwidth
        \advance\leftmargin\labelsep\usecounter{enumi}}}
 \relax
\catcode`\@=11
\def\marginnote#1{}
\newcount\hour
\newcount\minute
\newtoks\amorpm
\hour=\time\divide\hour by60
\minute=\time{\multiply\hour by60 \global\advance\minute by-
\hour}
\edef\standardtime{{\ifnum\hour<12 \global\amorpm={am}%
    \else\global\amorpm={pm}\advance\hour by-12 \fi
    \ifnum\hour=0 \hour=12 \fi
    \number\hour:\ifnum\minute<100\fi\number\minute\the\amorpm}}
\edef\militarytime{\number\hour:\ifnum\minute<100\fi\number\minute}

\def\draftlabel#1{{\@bsphack\if@filesw {\let\thepage\relax
  \xdef\@gtempa{\write\@auxout{\string
    \newlabel{#1}{{\@currentlabel}{\thepage}}}}}\@gtempa
    \if@nobreak \ifvmode\nobreak\fi\fi\fi\@esphack}
     \gdef\@eqnlabel{#1}}
\def\@eqnlabel{}
\def\@vacuum{}
\def\draftmarginnote#1{\marginpar{\raggedright\scriptsize\tt#1}}
\def\draft{\oddsidemargin -.5truein
        \def\@oddfoot{\sl preliminary draft \hfil
        \rm\thepage\hfil\sl\today\quad\militarytime}
        \let\@evenfoot\@oddfoot \overfullrule 3pt
        \let\label=\draftlabel
        \let\marginnote=\draftmarginnote
\def\@eqnnum{(\theequation)\rlap{\kern\marginparsep\tt\@eqnlabel}%
\global\let\@eqnlabel\@vacuum}  }
\def\preprint{\twocolumn\sloppy\flushbottom\parindent 1em
        \leftmargini 2em\leftmarginv .5em\leftmarginvi .5em
        \oddsidemargin -.5in    \evensidemargin -.5in
        \columnsep 15mm \footheight 0pt
        \textwidth 250mmin      \topmargin  -.4in
        \headheight 12pt \topskip .4in
        \textheight 175mm
        \footskip 0pt
\def\@oddhead{\thepage\hfil\addtocounter{page}{1}\thepage}
        \let\@evenhead\@oddhead \def\@oddfoot{} \def\@evenfoot{}
}
\def\titlepage{\@restonecolfalse\if@twocolumn\@restonecoltrue\onecolumn
     \else \newpage \fi \thispagestyle{empty}\c@page\z@
        \def\thefootnote{\fnsymbol{footnote}} }
\def\endtitlepage{\if@restonecol\twocolumn \else  \fi
        \def\thefootnote{\arabic{footnote}}
        \setcounter{footnote}{0}}  
\catcode`@=12
\relax

\def\ps@headings{\def\@oddfoot{}\def\@evenfoot{}
\def\@oddhead{\hbox{}\hfill
        \makebox[.5\textwidth]{\raggedright\ignorespaces --\thepage{}--
        \hfill }}
\def\@evenhead{\@oddhead}
\def\subsectionmark##1{\markboth{##1}{}}
}
\ps@headings
\relax
\newcommand{\newc}{\newcommand}

\newc{\ra}{\rightarrow}
\newc{\lra}{\leftrightarrow}
\newc{\beq}{\begin{equation}}
\newc{\be}{\begin{equation}}
\newc{\eeq}{\end{equation}}
\newc{\ee}{\end{equation}}
\newc{\bea}{\begin{eqnarray}}
\newc{\eea}{\end{eqnarray}}
\def\eps{\epsilon}

\newc{\ba}{\begin{eqnarray}}
 \newc{\ea}{\end{eqnarray}}
\newcommand{\e}{\varepsilon}
\newcommand{\eb}{\bar\varepsilon}
\renewcommand{\l}{\lambda}
\newcommand{\lb}{\bar\lambda}
\begin{document}
\def\firstpage#1#2#3#4#5#6{
\begin{titlepage}
\nopagebreak
\title{\begin{flushright}
        \vspace*{-0.8in}
{ \normalsize  hep-ph/9909206 \\
CERN-TH/99-268 \\
  IOA-18/1999 \\
August 1999 \\
}
\end{flushright}
\vfill
{#3}}
\author{\large #4 \\[1.0cm] #5}
\maketitle
\vskip -7mm
\nopagebreak
\begin{abstract}
{\noindent #6}
\end{abstract}
\vfill
\begin{flushleft}
\rule{16.1cm}{0.2mm}\\[-3mm]

\end{flushleft}
\thispagestyle{empty}
\end{titlepage}}

\def\simlt{\stackrel{<}{{}_\sim}}
\def\simgt{\stackrel{>}{{}_\sim}}
\date{}
\firstpage{3118}{IC/95/34}
{\large\bf New fermion  mass textures from  anomalous $U(1)$ symmetries\\
with baryon  and lepton number conservation}
{G.K. Leontaris$^{\,a,b}$
and J. Rizos$^{\,b}$}
{\normalsize\sl
$^a$Theory Division, CERN, CH 1211 Geneva 23, Switzerland\\[2.5mm]
\normalsize\sl
$^b$Theoretical Physics Division, Ioannina University,
GR-45110 Ioannina, Greece\\[2.5mm]
 }
{In this paper, we present solutions to the fermion mass hierarchy problem
in the context  of the minimal supersymmetric standard theory augmented by
an anomalous family--dependent $U(1)_X$ symmetry. The latter is spontaneously
broken  by non--zero  vevs of a pair of singlet fields whose magnitude is 
determined through the $D$-- and $F$--flatness conditions of the superpotential.
 We derive the general
solutions to the anomaly cancellation conditions and show that they allow numerous
choices for the $U(1)_X$ fermion charges which give several fermion mass textures in
agreement with the observed fermion mass hierarchy and mixing. Solutions with $U(1)_X$
fermion charge assignments are found which forbid or substantially suppress the
dangerous baryon and lepton number violating operators and the lepton--higgs mixing
coupling while a higgs mixing mass parameter ($\mu$--term) can be fixed at the
electroweak level. We give a general classification of the fermion mass textures with
respect to the sum of the doublet--higgs $U(1)_X$--charges and show that suppression
of dimension--five operators  naturally occurs for various charge
 assignments. We work out cases which
retain a quartic term providing the left--handed neutrinos with Majorana masses in the
absence of right--handed neutrino components and consistent with the experimental
bounds. Although there exist solutions which naturally combine all the above features
with rather natural $U(1)_X$ charges, the suppression of the $\mu$--term occurs for
particular assignments.}

\newpage
\section{Introduction}

The minimal supersymmetric extension of the standard model theory (MSSM) has had a
remarkable success in explaining the low energy parameters in the context of
unification scenario. Among them, the measured values of the strong coupling constant
$\alpha_s(m_W)$ and the weak mixing angle $\sin^2\theta_W(m_W)$ are in perfect
agreement with those predicted when the unification scale is taken to be of the order
$M_U\sim 10^{16}$GeV and the
 only contribution of the MSSM spectrum is assumed to the
beta-function coefficients for the gauge coupling running. These remarkable properties
of the simplest unified model, naturally raise the question whether the fermion mass
spectrum observed in low energies is also reproduced from few basic symmetry
principles encountered
 at the unification scale.

The experience from string model building has shown that a natural step towards
this simplification is to assume the existence of $U(1)$ symmetries which
distinguish the various families. A further indication that additional symmetries
beyond the standard gauge group exist, has been the observation that the fermion
mixing angles and  masses have values consistent with the appearance of
``texture'' zeros in the mass matrices~\cite{Ramond:1993kv}. More precisely, it
has  been observed that in string model building one usually ends up with the
effective field theory model  which, in addition to the non--abelian gauge group
includes an anomalous abelian gauge symmetry whose anomaly is cancelled by the
Green--Schwarz mechanism~\cite{Green:1984sg}. In fact, this mechanism allows for
the existence of a gauged $U(1)_X$ whose anomaly is cancelled by assigning a
non--trivial transformation to an axion which couples universally to all gauge
groups. In the spectrum of a string model, there are usually singlet fields
$\phi_i,\,\bar\phi_i$ charged under this $U(1)$ symmetry which develop vacuum expectation
values (vevs) in order to satisfy the $F$-- and
$D$--flatness conditions of the superpotential. This results to a spontaneous
breaking of the anomalous $U(1)$ symmetry, naturally at some scale one order of
magnitude less than the string (unification) scale.

 Surprisingly, the existence of an anomalous or non--anomalous $U(1)$   symmetry
has remarkable implications in low energy physics: for example, one may try to
explain~\cite{Froggatt:1979nt,Ibanez:1993fy,Ibanez:1994ig,Leurer:1994gy,
Binetruy:1995ru,Nir,Dreiner:1995ra,Dudas:1995yu,Altarelli:1998ns} the mass
hierarchies observed in the quark and charged leptonic sector. In this approach,
all quark, lepton and higgs fields are charged under the extra abelian symmetry.
The charges are chosen so that when the $U(1)$ symmetry is unbroken, only the
third generation is massive and all mixing angles are zero. However, when the
singlet fields obtain a non--zero vev, symmetry breaking terms gradually fill in
the fermion mass matrices and generate a hierarchy of mass scales and mixing angles.
It turns out that  the symmetry breaking terms appearing in the fermion mass
matrices may be expressed as powers of a few expansion parameters leading to a
rather impressive predictability of the whole scheme.  If further the $U(1)$ is
anomalous, then the vacuum expectation values of the singlets are also given in
terms of the unification (string)  scale and definite predictions may arise for
the masses and mixing angles. In fact, it will turn out that successful
hierarchical mass patterns appear only if the $U(1)_X$ symmetry is anomalous.

It is rather interesting that this scenario may also give the correct prediction
for the weak mixing angle without assuming unification. It was shown in
Ref.~\cite{Ibanez:1993fy} that in the presence of an anomalous $U(1)_X$ symmetry, the
value of $\sin^2\theta_W$ could be predicted in terms of the
$U(1)_X$--charges of the massless fermions. The anomaly cancellation mechanism
may work in a way that the gauge couplings have the correct predictions in low
energies. The simplest possibility of symmetric mass matrices was worked out in
Ref.~\cite{Ibanez:1994ig} and found that the hierarchical pattern of the fermion mass
spectrum can be successfully reproduced.

In the present paper, we wish to extend the analysis by considering the most general
$U(1)_X$ symmetry with fermion charges respecting the anomaly cancellation conditions.
We first find that, even in the simple case of a $U(1)$ factor obtained from linear
combination of the standard model symmetries, the fermion mass matrix structure is
richer than that exhibited in\cite{Ibanez:1994ig}. This is a simple consequence of the
fact that the vacuum expectation values of the two singlet fields $\phi$ and
$\bar\phi$ differ from each other since they have to respect the $D$--term anomaly
cancellation condition. Going further, we find that a general family--dependent
$U(1)$ anomalous symmetry generates four approximate texture--zero mass matrices
of Table 1. It is found that, the higgs $U(1)_X$ charges play a crucial role,
particularly in the determination of the lepton textures as well as the baryon
and lepton number violation. For a non--zero sum of the $U(1)_X$ higgs charges
it is possible to ban all dangerous dimension five proton decay operators. We
further find that one can choose a consistent set of $U(1)_X$ charges which
prevent the appearance of an unacceptably large Majorana mass term for the
left--handed neutrino.

The paper is organised as follows: In Section 2, we introduce the notation for fermion
charges, we set our assumptions and solve the constraints from mixed anomalies for the
fermion and higgs $U(1)_X$ charges. In Section 3 we derive the general forms of the
quark and lepton mass matrices. Using specific values for the charges, consistent with
the solutions obtained in Section 2, we classify with respect to the sum of the
higgs--doublet $ U(1)_X$ charges all possible fermion mass textures. In Section 4 we
analyse the baryon and lepton number violating operators, as well as other dangerous
terms which are not prevented from the standard model gauge symmetry. We impose
constraints to eliminate these dangerous operators and in Section 5 we present 
particular examples of fermion matrices which fulfill these requirements. In Section 6
we present our conclusions.
\begin{table}
\centering
\begin{tabular}{|c|c|c|} \hline
Texture & $M_U$ & $M_D$ \\ \hline
$T_1$ & $\left(
\begin{array}{ccc}

0 & \sqrt{2}\lambda^6 & 0 \\
\sqrt{2}\lambda^6 & \lambda^4 & 0 \\
0 & 0 & 1
\end{array}
\right)$ & $\left(
\begin{array}{ccc}

0 & 2\lambda^4 & 0 \\
2\lambda^4 & 2\lambda^3 & 4\lambda^3 \\
0 & 4\lambda^3 & 1
\end{array}
\right)$
\\ \hline

$T_2$ & $\left(
\begin{array}{ccc}

0 & \lambda^6 & 0 \\
\lambda^6 & 0 & \lambda^2 \\
0 & \lambda^2 & 1
\end{array}
\right)$ & $\left(
\begin{array}{ccc}

0 & 2\lambda^4 & 0 \\
2\lambda^4 & 2\lambda^3 & 2\lambda^3 \\
0 & 2\lambda^3 & 1
\end{array}
\right)$
\\ \hline

$T_3$ & $\left(
\begin{array}{ccc}

0 & 0 & \sqrt{2}\lambda^4 \\
0 & \lambda^4 & 0 \\
\sqrt{2}\lambda^4 & 0 & 1
\end{array}
\right)$ & $\left(
\begin{array}{ccc}

0 & 2\lambda^4 & 0 \\
2\lambda^4 & 2\lambda^3 & 4\lambda^3 \\
0 & 4\lambda^3 & 1
\end{array}
\right)$
\\ \hline

$T_4$ & $\left(
\begin{array}{ccc}

0 & \sqrt{2}\lambda^6 & 0 \\
\sqrt{2}\lambda^6 & \sqrt{3}\lambda^4 & \lambda^2 \\
0 & \lambda^2 & 1
\end{array}
\right)$ & $\left(
\begin{array}{ccc}

0 & 2\lambda^4 & 0 \\
2\lambda^4 & 2\lambda^3 & 0 \\
0 & 0 & 1
\end{array}
\right)$
\\ \hline

$T_5$ & $\left(
\begin{array}{ccc}

0 & 0 & \lambda^4 \\
0 & \sqrt{2}\lambda^4 & \frac{\lambda^2}{\sqrt{2}} \\
\lambda^4 & \frac{\lambda^2}{\sqrt{2}} & 1
\end{array}
\right)$ & $\left(
\begin{array}{ccc}
0 & 2\lambda^4 & 0 \\
2\lambda^4 & 2\lambda^3 & 0 \\
0 & 0 & 1
\end{array}
\right)$
\\ \hline
\end{tabular}
\caption{The five symmetric  texture--zero mass matrices  for
the up-- and down-- quarks  consistent with the observed hierarchical
 pattern\cite{Ramond:1993kv}. }
\label{textures}
\end{table}

\section{\label{secc}The Solutions to Anomaly Cancellation Conditions}

The Yukawa terms of the superpotential needed to provide masses to quarks and
leptons are $SU(3)\times SU(2)_L\times U(1)_Y$ gauge invariant and are written
as follows
\ba
{\cal W} &=& \lambda_{ij}^u Q_i U^c_j H_2 + \lambda_{ij}^dQ_i D^c_j H_1
               +\lambda_{ij}^e L_{i} E^c_j H_1.
\label{sup1}
\ea
Although all these terms are invariant under the standard model gauge group, there is
no explanation why some of the Yukawa couplings are required to be much smaller than
others to account for the fermion mass hierarchy. Extending the standard
gauge group by one anomalous $U(1)_X$ with the MSSM fields being charged under this
abelian factor, only terms which are invariant under this symmetry remain in the
superpotential. The observed low energy hierarchy of the fermion mass spectrum and the
demand to have natural values of the Yukawa couplings
$\lambda_{ij}$ of order one, suggest that only tree--level couplings associated
with the third generation should remain invariant. In this case, the mass terms
involving some of the lighter fermions are generated through non--renormalizable
superpotential couplings at some order. These higher order invariants are formed
by adding to the tree--level coupling an appropriate number of singlet fields
($\bar\phi$ or $\phi$) which  compensate the excess of the $U(1)_X$ charge. Since
the anomaly cancellation mechanism \cite{DSW} requires vevs for the singlet fields
that are about an order of magnitude below the unification scale, the above
scenario naturally reproduces hierarchical fermion mass  spectra. We therefore
assume that the anomalous charge  of the singlet fields $\phi$ and $\bar\phi$
is $+1,-1$ respectively  which is equivalent to measuring all charges in $\phi$ charge
units.

 In the present work we are interested in  symmetric fermion mass matrices.
To obtain a symmetric structure we need to define proper constraints on the
fermion charges under the $U(1)_X$ symmetry.  Under the assignments of Table
\ref{uxa} the charges of the mass matrices are $C^U_{ij}=q_i+u_j$,
$C^D_{ij}=q_i+d_j$ and $C^E_{ij}=\ell_i+e_j$.
\begin{table}
\begin{center}
\begin{tabular}{cc}
Field&Charge\\
$Q_i$&$q_i$\\
$D^c_i$&$d_i$\\
$U^c_i$&$u_i$\\
$L_i$&$\ell_i$\\
$E^c_i$&$e_i$\\
$H_2$&$h_2$\\
$H_1$&$h_1$\\
\end{tabular}
\end{center}
\caption{\label{uxa}Charge assignments for MSSM fields under the  $U(1)_X$.}
\end{table}
The conditions for symmetric mass matrices in the above notation take the form
\ba
q_i+u_j&=&q_j+u_i\nonumber\\
q_i+d_j&=&q_j+d_i\label{symcon}\\
\ell_i+e_j&=&\ell_j+e_i.\nonumber
\ea
The requirement that the third generation has tree--level couplings imposes
the constraints,
\ba
q_3+u_3+h_2&=&0\nonumber\\
q_3+d_3+h_1&=&0\label{treecon}\\
\ell_3+e_3+h_1&=&0.\nonumber
\ea
After imposing the conditions (\ref{symcon}),(\ref{treecon}) the charges of
the possible quark couplings to the appropriate higgs field take the form,
\be
C^{QU^c H_2}=C^{QD^cH_1}=\left(\begin{array}{ccc}
2(q_1-q_3)&(q_1-q_3)+(q_2-q_3)&q_1-q_3\\
(q_1-q_3)&2(q_2-q_3)&q_2-q_3\\
(q_1-q_3)&(q_2-q_3)&0\\
\end{array}\right).\label{qcharge}
\ee
Similarly, for the charged leptons we have,
\be
C^{LE^cH_1}=\left(\begin{array}{ccc}
2(\ell_1-\ell_3)&(\ell_1-\ell_3)+(\ell_2-\ell_3)&\ell_1-\ell_3\\
(\ell_1-\ell_3)&2(\ell_2-\ell_3)&\ell_2-\ell_3\\
(\ell_1-\ell_3)&(\ell_2-\ell_3)&0\\
\end{array}\right).\label{lcharge}
\ee
We observe that the charges of the up and down quark entries are the same.
This  result is obtained only due to the fact that we require symmetric
textures and one--tree level coupling for each one of the quark matrices.
Further, the quark  charge--entries depend only on two combinations, $q_1-q_3$
and $q_2-q_3$.  This is also the case for the leptons,  with the replacements
$q_i\ra \ell_i$. Anomaly cancellation conditions will give  further relations
between $q_i$ and $\ell_i$ charges, so if $U(1)_X$ charges are  somehow fixed
in the quark sector, then one ends up with predictions in the lepton
matrices.

We proceed with the analysis of the quarks. At this stage, one can readily
conclude  that in order to have acceptable quark masses we must have
\be
q_1-q_3=\frac{n}{2}\ , \  q_2-q_3 =\frac{m}{2}\ \ \mbox{\rm where}\
m+n\ne0,\ m,n=\pm1,\pm2,\dots\label{qi}
\ee
We do not  write down a similar parametrization for the charged lepton entries
since, as we will see, due to the various conditions we have imposed we will be
able to express them in terms of the quark entries.   We will discuss in detail
the remaining constraints on the quark and lepton entries in the subsequent section
but first we need to
deal with the mixed anomalies associated with the MSSM and $U(1)_X$ gauge groups.

It is well known that the MSSM is anomaly free. The introduction of an extra anomalous
$U(1)_X$ group factor leads to anomalies which should be absorbed. As already
discussed, the Green--Schwarz anomaly cancellation mechanism may cancel the pure
$U(1)_X$ anomaly and mixed $U(1)_X$--gravitational anomalies, however there are mixed anomalies of the form $A_i=\left(G_i G_i
U(1)_X\right)$ where $G_i=(SU(3),SU(2),U(1)_Y)$. In terms of the $U(1)_X$ charges
these are written as
\ba
 A_3:&&\sum q_i+ \frac 12 \sum (u_i + d_i) \label{a3}\\
A_2:&&\frac 32 \sum q_i +\frac 12\sum \ell_i + \frac 12 (h_1+ h_2)
\label{a2}\\
A_1:&&\frac 16 \sum q_i +\frac 13 \sum d_i +
\frac 43 \sum u_i + \frac 12 \sum \ell_i
+\sum e_i + \frac 12 (h_1+h_2).
\label{a1}
\ea

It was pointed out\cite{Ibanez:1993fy} that in a model where the anomalies are
canceled  through the Green--Schwarz mechanism, the mixed anomalies with the
standard model gauge group are proportional to the corresponding  Kac--Moody level,
\ba
\frac{A_3}{A_2}&=&\frac{k_3}{k_2}\label{ac1}\\
\frac{A_2}{A_1}&=&\frac{k_2}{k_1}\,.\label{ac2}
\ea
There are also mixed anomalies of the form $A_0=\left(U(1)_Y  U(1)^2_X\right)$
which should be zero:
\be
A_0:\ \sum q_i^2 +\sum d_i^2-2\sum u_i^2-\sum \ell_i^2+\sum e_i^2+(h_2^2-h_1^2)=0
\label{a0}.
\ee

We should note here that in the calculation of anomalies we have only considered
the fields of the MSSM spectrum and a pair of singlets $\phi, \bar\phi$ which
are necessary to break the $U(1)_X$ symmetry and create the non--renormalizable
terms which fill in the fermion mass matrices. In string constructions however,
there are additional particles (some of them carry fractional charges), which
may also be charged under $U(1)_X$. Even in the case that these fields belong
to non--trivial representations of $SU(3)\times SU(2)\times U(1)_Y$ group they
usually come in pairs with opposite $U(1)_X$ charges so they do not contribute
to the mixed anomalies.

The conditions (\ref{ac1},\ref{ac2}) have rather remarkable implications
on the determination of the low energy parameters. Indeed, to confront with
the standard unification scenario we have to impose the conditions,
\ba
\sin^2\theta_W(M_U)&=&\frac{3}{8} \\
k_3&=&k_2
\ea
which lead to the constraints,
\be
\frac{A_3}{A_1}=\frac{A_2}{A_1}=\frac{3}{5}\,.\label{ac}
\ee
To proceed further, we should combine the equations obtained from the anomaly
cancellation with the symmetry constraints (\ref{symcon}). These can be solved with
respect to the charges  $u_1,u_2,d_1,d_2,e_1$ and  $e_2$ 
while the constraints (\ref{treecon}) can be
solved with respect to $u_3,d_3$ and $e_3$. Furthermore, we may solve the constraints
(\ref{ac}) with respect to the charges
$q_3, \ell_3$, while treating  as parameters the sums  $h_+$, $q_+$, $\ell_+$:
\ba
h_+=h_1+h_2,\,\, q_+=q_1+q_2,\,\, \ell_+=\ell_1+\ell_2.
\label{rep}
\ea
 Then, in terms of (\ref{rep}) we obtain,
\ba
\ell_3&=& \frac 1{48}\left[-5 h_+ - 18 (q_+-\ell_+)\right]
\label{la3}
\\
q_3&=& \frac 1{48}\left[-17 h_+ + 6 (q_+-\ell_+)\right]
\label{qu3}
\ea
 This parametrization  simplifies considerably the analysis of the quadratic
constraint (\ref{a0}). Indeed, substituting the above solutions into (\ref{a0})
we arrive  to the equation,
\ba
   6 h_+^2 + 5 ( 11 q_+ + \ell_+  - 4  h_2) h_+
   - 6 ( \ell_+ + 3 q_+) (q_+ - \ell_+ + 4 h_2) = 0
   \label{eqh+}
\ea
which can be solved easily.  We find it convenient now to solve the above
equation for $h_2$, keeping as parameters the sums  $h_+$, $q_+$, $\ell_+$
as previously. We find two  solutions depending on the value of $h_+$:
\begin{itemize}
\item
For $h_+=2(q_{+}-2q_3)\ne0$ we obtain the simple relations,
\ba
\frac{3}{16}\,\ell_{+}=\frac{1}{3}\,q_3=\frac{3}{8}\,\ell_3=q_{+}\, .
\label{sol0}
\ea
\item
For $h_{+}\ne2(q_{+}-2q_3)$ we obtain the solution,
\ba
h_2&=&\frac{-5\,h_{+}^2+h_{+}(q_{+}-29\,q_3)+24q_3(q_{+}-2q_3)}{6(h_{+}-
2q_{+}+4q_3)} \nonumber\\
\ell_{+}&=&q_{+}-8q_3-\frac{17}{6}h_{+}\label{h2eq}\\
\ell_3&=&-\frac{7}{6}h_{+}-3q_3.\nonumber
\ea
\end{itemize}
The first solution is characterized by three parameters $q_1,q_2, h_2$ or
equivalently  $m,n,h_2$  (see Eq.(\ref{qi})). The second is a four parameter
solution that  depends on $m,n,q_3,h_+$ and  $\ell_2$. As it will become clear
later (see Section 3), $\ell_2$ can be exchanged with an integer
parameter $k$ and $h_+$ has to be integer too. Thus the first solution depends
on two  integer and one non--integer parameters while the second depends one
three integers and  one non--integer ones. The detailed charge assignments
for each solution are  shown in Tables  \ref{tsol0},\ref{tsol1} respectively.

We wish to emphasize here that, these are the most general solutions to the
anomaly  cancellation conditions under the assumptions of symmetric mass
matrices and  tree--level Yukawa couplings for the third generation with
$b-\tau$ unification. Using  the parametrization (\ref{rep}), we have
succeeded to linearize the quadratic equation  constraint and express the
solutions in terms of a few parameters. The solutions above  are suggestive
for a classification with respect to the sum $h_+$ of the higgs charges.

\begin{table}
\centering
\begin{tabular}{|l|c|c|c|}
\hline
field&\multicolumn{3}{c|}{generation}\\
\hline
&1&2&3\\
\hline
$Q$&$\frac{2n-3m}{10}$&$\frac{2m-3n}{10}$&$-\frac{3(m+n)}{10}$\\
$U^c$&$\frac{3m+8n}{10}-h_2$&$\frac{8m+3n}{10}-h_2$&$\frac{3(m+n)}{10}-h_2$\\
$D^c$&$-\frac{7m+2n}{10}+h_2$&$-\frac{7n+2m}{10}+h_2$&$-\frac{7(n+m)}{10}+h_2$\\
$L$&$-\frac{15k+23m+8n}{30}$&$\frac{15k+7m-8n}{30}$&$-\frac{4(m+n)}{15}$\\
$E^c$&$-\frac{15k+37m+22n}{30}+
h_2$&$\frac{15k-7m-22n}{30}-h_2$&$\frac{11(m+n)}{15}-h_2$\\
\hline
\multicolumn{4}{|c|}{Higgs}\\
\hline
$H_1$&$m+n-h_2$&$H_2$&$h_2$\\
\hline
\end{tabular}
\caption{$U(1)_X$ charge assignments for the case $h_{+}=m+n$ (Equation (\ref{sol0})).}
\label{tsol0}
\end{table}
\begin{table}
\centering
\begin{tabular}{|l|c|c|c|}
\hline
field&\multicolumn{3}{c|}{generation}\\
\hline
&1&2&3\\
\hline
$Q$&$\frac{n}{2}+q_3$&$\frac{m}{2}+q_3$&$q_3$\\
$U^c$&$\frac{n}{2}-q_3-h_2$&$\frac{m}{2}-q_3-h_2$&$-q_3-h_2$\\
$D^c$&$\frac{n}{2}-q_3-h_++h_2$&$\frac{m}{2}-q_3-h_++h_2$&$-h_++h_2-q_3$\\
$L$&$\frac{n-k}{2}-3q_3-5\frac{h_+}{3}$&
$\frac{m+k}{2}-3q_3-7\frac{h_+}{6}$&$-3q_3-7\frac{h_+}{6}$\\
$E^c$&$\frac{n-k}{2}+3q_3-\frac{h_+}{3}+
h_2$&$\frac{m+k}{2}+3q_3+\frac{h_+}{6}+h_2$&$\frac{h_+}{6}+h_2+3 q_3$\\
\hline
\multicolumn{4}{|c|}{Higgs}\\
\hline
$H_1$&$h_+-h_2$&$H_2$&$h_2$\\
\hline
\end{tabular}
\caption{$U(1)_X$ charge assignments for the case $h_{+}\ne m+n$. The value of
$h_2$ is not an independent parameter but it is related to $q_3,h_+,m,n$ through
Eq. \ref{h2eq}. The integer $k$ is defined as $k=2\ell_2+6q_3-m+\frac 73 h_+$ (see
Section \ref{seck}).}
\label{tsol1}
\end{table}
At this point, we find it useful to close this section with a remark on the
necessity of the $U(1)$--symmetry of being anomalous. Indeed, one may wonder
whether an anomaly free abelian symmetry can comply with the
above phenomenological requirements.
Actually one can easily derive the most general solution of the constraints
(\ref{symcon}), (\ref{treecon}) together with $A_3 = A_2 = A_1 = A_0 =0$ that
gives also vanishing trace for $U(1)_X$ and vanishing $U(1)_X^3$ anomaly. This
solution is  $2q_3 = d_3 = -u_3/2 = -\ell_+/3 = e_3/3=q_+$, $u_1=-3 q_+/2-q_2$,
$u_2 = -5q_+/2 + q_2, d_2 = q_+/2 + q_2$, $d_1=3q_+/2-q_2$, $\ell_1=-3q_+-\ell_2$,
$e_1=3q_+/2-\ell_2$, $e_2=9q_+/2+\ell_2$, $ h_1=-h_2=l_3=-3 q_+/2$.  Unfortunately
this solution predicts two additional tree--level couplings, namely  $Q_2 U_1
H_2$  and $Q_1 U_2 H_2$ which are not consistent with phenomenological
requirements (see Table \ref{textures}). Thus one is forced to search for an
anomalous $U(1)$ symmetry.

\section{\label{seck} The Derivation of the Fermion Mass Matrices}

In the subsequent analysis with regard to the derivation of the fermion mass
matrices and the investigation of baryon and lepton number violation,
a crucial role is  played by the value of the parameter $h_+\equiv h_1+h_2$.
Some first conclusions may  be drawn by inspection of the forms of the charge
matrices (\ref{qcharge},\ref{lcharge}), when they are written in terms of the
parameters $m,n,h_+$. Thus, the up and down quark mass matrices are independent
of  the value of $h_+$, while, on the contrary, the structure of the charged
lepton mass matrix  depends decisively on $h_+$. It can be also easily checked
that the baryon and lepton number  violating operators are $h_+$--dependent. It
is therefore convenient for our subsequent  analysis to distinguish three cases
for the $h_+$ value: we will first examine the  case $h_+=0$ which means that the
two higgs doublets possess opposite charges. Next we consider $h_+$ to be an integer
and finally we comment on the non--integer values of $h_+$.

\subsection{$\,\,h_+=0$}
Starting from this particular value $h_+$ , we first find that one of the
two solutions (\ref{sol0},\ref{h2eq}) does not lead to sensible results.
Indeed, by a simple inspection  we infer that solution (\ref{eqh+}) would
imply $m+n=0$ and therefore, a tree--level mass for the 12 and 21 entries.
Thus, there is only one solution for (\ref{eqh+}), namely (\ref{h2eq}) which
for $h_+=0$ takes
the form,
\ba
 \ell_1+\ell_2=(q_1+q_2)-8q_3\ ,
\ \ell_3=-3q_3\ ,\ h_2=-h_1=-2q_3\, .
\ea
We can easily calculate the charges which are given in Table \ref{thp0} in
terms of the four free--parameters.

Using these charge assignments one finds that the quark-- and lepton--charge
matrices take the form
\ba
C^{QU^cH_2}=C^{QD^cH_1}&=&\left(\begin{array}{ccc}
n&\frac{m+n}{2}&\frac{n}{2}\\
\frac{m+n}{2}&{m}&\frac{m}{2}\\
\frac{n}{2}&\frac{m}{2}&0\\
\end{array}\right)\label{qna}
\\
C^{LE^cH_1}&=&\left(\begin{array}{ccc}
n-k&\frac{m+n}{2}&\frac{n-k}{2}\\
\frac{m+n}{2}&m+k&\frac{m+k}{2}\\
\frac{n-k}{2}&\frac{m+k}{2}&0\\
\end{array}\right)
\label{lep}
\ea
where $m,n$  are given in (\ref{qi})  while all the additional dependence
of the lepton matrix has been absorbed in the parameter $k$  defined as
follows
\ba
 k&=&2\ell_2+6 q_3-m\,.
\label{nmk}
\ea

Let us now come to the parameters entering the quark and charged lepton mass
matrices.  We first note that the only tree--level couplings entering the
fermion mass textures  are those corresponding to the $33$--entries of
(\ref{qna}) and (\ref{lep}), i.e., the  third generation mass terms for the
up, down and lepton fields, i.e., $Q_3 t^c H_2$, $Q_3 b^c  H_1$ and
$L_3 \tau^c H_1$.  The remaining mass matrix entries are expected to be
generated from non--renormalizable terms formed by proper powers of the
singlet  fields $\phi/M_1$, $\bar\phi/M_1$ and $\phi/M_2$, $\bar\phi/M_2$.
The powers of  non--renormalizable terms have to be such that the charges of
the entries in the  matrices (\ref{qna}),(\ref{lep}) are cancelled out. The
singlets are divided by the  mass parameters $M_1, M_2$ which refer to some
high energy scales. If the dominant  source of these terms is from string
compactification, then $M_1 = M_2 \equiv M $ and  there are only two expansion
parameters which enter in the mass matrices, namely
$\phi/M$ and $\bar\phi/M$. It is also possible that additional vector--like
higgs pairs  may acquire their mass via spontaneous breaking after
compactification. Then, a strong  violation of the $SU(2)_R$ symmetry of the
quark sector may occur, and as a result
$M_1\ne M_2$~\cite{Ibanez:1994ig}.

We further point out here that in the case of an anomalous $U(1)$ symmetry
we should  necessarily take $\phi\ne \bar\phi$. This is because the cancellation
of the $D$--term requires these values to differ from each other. In particular,
the Green--Schwarz anomaly cancellation mechanism generates a constant
Fayet--Iliopoulos\cite{FI} contribution to the $D$--term of the anomalous
$U(1)_X$.  This is proportional to the trace of the anomalous charge over all
fields capable of obtaining non--zero vevs. To preserve supersymmetry the
following
$D$--flatness condition should be satisfied\cite{DSW,ADS},
\ba
  \sum_{\phi_j} Q_j^X |\phi_j|^2&=&-\xi \ne 0
\label{Deq}
\ea
where $Q_J^X$ is the $U(1)_X$ charge of the field $\phi_j$,  while the sum extends
over all possible singlet fields and the parameter $\xi$ is proportional to the trace
of the anomalous  $U(1)_X$.  Clearly, in the case of two singlets with opposite
charges --as in our case-- one should have
\ba
|\langle\phi\rangle|^2 - |\langle\bar\phi\rangle|^2 = -\xi
\label{acc}
\ea
Thus, in the construction of the quark and lepton mass matrices
in the general case we may define the following four  parameters
\ba
\e=\frac{\phi}{M_1},&  \eb=\displaystyle\frac{{\displaystyle\bar\phi}}{M_1}\\
\l=\frac{\phi}{M_2},&  \lb=\displaystyle\frac{{\bar\phi}}{M_2}.
\ea
According to our previous discussion, the  parameters $\e$
and $\eb$ should appear in the up--quark mass matrix while the
second set, i.e. $\l$ and $\lb$, in the down quark and charged
lepton mass matrices. Thus the possible mass terms should
have one of the following forms
\ba
Q_i U_j H_2 \e^{r_{ij}},& Q_i U_j H_2 \eb^{r_{ij}}
\nonumber
\ea
for the up quarks and,
\ba
Q_i D_j^c H_1 \e^{s_{ij}},& Q_i D_j^c H_2 \eb^{s_{ij}}
\nonumber\\
L_i E_j^c H_1 \e^{p_{ij}},& L_i E_j^c H_1 \eb^{p_{ij}}
\nonumber
\ea
for the down quarks and charged leptons. Here, $r_{ij}, s_{ij}$ and $p_{ij}$ are
numbers which represent the necessary powers of the expansion parameters in order to
cancel the charge of the corresponding $ij$--entry in (\ref{qna}) and (\ref{lep}).
Clearly these numbers have to be integers.

In the convenient parametrization (\ref{nmk})  the charge entries depend only
on the three free parameters  $m,n,k$. Every charge entry is scaled with
the charges of the singlet fields $\phi,\bar\phi$. Thus, without loss of
 generality we may simply take the charge  of the latter to be $+1$ and $-1$
respectively. Under a certain choice of the $U(1)_X$ charges of the various
fermion and higgs fields, the entries of the charge--matrices (\ref{qna}) and
(\ref{lep}) can be either positive or negative. (We exclude the case of charges
leading to zeros in these entries since this would lead to an additional
tree--level order entry in the mass matrix). Now, if the charge of an entry is
positive we may cancel this only by adding powers of the singlet fields carrying
negative $U(1)_X$ charge. On the contrary, entries with negative charge require
powers of  positively charged singlets. Therefore, in certain choices it is
possible that two expansion parameters may enter in the mass matrices, leading
thus to new structures. Note the fact that, there is no loss of predictability
compared to previous cases\cite{Ibanez:1994ig} where only one expansion parameter
was used in the fermion mass textures. Indeed, the two singlet vevs are related
through the equation (\ref{Deq}), while the parameter $\xi$ in the right--hand
side of this equation is completely determined by the trace of the anomalous
symmetry, the value of the common gauge coupling at $M_U$ and the string
(unification) scale itself\cite{DSW,ADS}.

In order for a mass entry to be generated, the specific combination  of the
free parameters $m,n,k$  entering the charge entry has to be integer.
Otherwise the corresponding   mass  entry is zero since no power of singlet
vevs with  $\pm 1$ charge could make the relevant Yukawa coupling invariant
under this $U(1)$. With the above remarks in mind, we are ready now to proceed
in the determination of the viable fermion mass textures.

{\bf \underline{Quarks}}

We proceed now with the determination of all  possible structures of
 the quark matrix. Although in this section we deal only with the case
$h_+ = 0$, we will soon see that even in the most general case  where
$h_+\ne 0$ the form of the quark mass matrices does not change as  long as we
impose the symmetric mass textures conditions 
on the  $U(1)_X$--charges and the requirement of
one tree--level coupling for each fermion matrix\footnote{As a matter of fact,
this requirement on the  $U(1)_X$--charges only ensures  that at least one
entry admits a tree--level coupling. It does not exclude the appearance of more
than one tree--level couplings. Such solutions {\it do} appear and are excluded
for phenomenological reasons.}. Thus, since only the 33--entry is filled up by
a renormalizable coupling, clearly as long as the $U(1)_X$ remains unbroken
the two lighter generations remain massless. When the singlets $\phi,$
$\bar\phi$ develop  non--zero vevs along the $D$--flat direction, their magnitude 
are of the order  $\langle \phi \rangle \sim$ $\langle \bar\phi \rangle$ $
\sim \xi$ which is approximately one order lower than the  unification scale.
Then the  anomalous symmetry is broken and the remaining  fermion mass matrix
entries are filled up with mass terms suppressed by powers of the expansion
parameters. These powers  depend on the certain choice of
$U(1)_X$ charges.

Let us now determine the possible viable cases for the parameters $m,n$ which
enter the quark mass matrices. In order to have a non--zero value for the second
generation quarks it is evident from the structure of the charge matrix
(\ref{qna}) that the parameter $m$ has to be an integer. Only in this  case at
least one of the entries 22,
23 / 32, may survive. With similar reasoning, while taking also into
consideration  the necessity of the Cabbibo mixing, we may also conclude that
the parameter $n$ has to be an integer too. No constraint from mixing effects can
be imposed in the case of leptons, however, due to the fact that $m,n$ are
integers we have also to take $k$ to be integer otherwise we would end up with
two massless charged lepton states. Therefore there are four
possibilities among which we  distinguish three viable cases, namely  {\it i)
}$m,n$=even, {\it ii)} $m$=odd,  $n$=even  {\it iii)} $m$=odd, $n$=odd. (The
case $m=$even $n=$odd does not lead to acceptable mixings.) We consider these
cases separately and further, we work out certain choices of $m,n$ pairs which
lead to viable mass textures with reasonable values of the expansion parameters.
We note that due to our freedom to have two different expansion parameters and to
adjust order--one coefficients in the mass matrix entries, additional pairs of
$m,n$ values are also possible. They imply different values for the expansion
parameters but do not lead to different textures thus they are not elaborated
here.

{\bf case}$ \, i :\,\, n, m$ even.\\
{$ i_A$}:  We first start with positive $n, m$ values. In this case all quark
charge--entries in (\ref{qna}) are positive, so the lowest power of singlets
needed to cancel this charge involves only the singlet $\bar\phi$ to a proper
power. (Additional contributions involving pairs $(\phi\bar\phi)^v$ are always
possible but relatively suppressed.) Taking $n=2 m>0$ we obtain a quark mass
structure similar to the texture $T_5$ of Table 1. E.g. for $m=4,\,n=8$ we get
\be
m_U=\left(\begin{array}{ccc}
\eb^8&\eb^6&\eb^4\\
\eb^6&\eb^4&\eb^2\\
\eb^4&\eb^2&1\\
\end{array}\right) .
\label{upi1}
\ee
This matrix is not actually an exact texture--zero as the $T_5$ case, however,
one can observe that the entries replacing the zeros of texture $T_5$ are highly
suppressed here. Remarkably, this texture is also an outcome of the string
derived flipped $SU(5)$ model\cite{Leontaris:1993wp,Ellis:1997ni}.

The corresponding down quark matrix has the same form but in general involves
a different expansion parameter, namely $\lb$.  This gives the freedom to
adjust the two parameters so that the correct hierarchy and Cabbibo
mixing arise\footnote{Note, however, that the same expansion parameter $\lb$
enters in the lepton sector, so additional constraints will also come from
the charged lepton mass eigenvalues.}.
Thus, the matrix takes the form:
\be
m_D=\left(\begin{array}{ccc}
\lb^8&\lb^6&\lb^4\\
\lb^6&\lb^4&\lb^2\\
\lb^4&\lb^2&1\\
\end{array}\right).
\label{upi2}\ee
We note here that the mass entries in the above textures are accurate up to order one
coefficients which are not calculable in this approach. As far as we know the
calculation of the coefficients is only possible in string models. The remarkable
fact, however, in the present simple approach is that one does not need to introduce
unnaturally small Yukawa couplings to explain the huge ratios of mass eigenstates. The
present procedure tell us that the hierarchical pattern is just a simple consequence
of the $U(1)_X$ symmetry.

{\it $i_B$}:
 Next, we give an example where both parameters enter in the structure of the quark
mass matrices. Thus, taking one of the integers to be negative, we may obtain textures
with $\e$ and $\eb$ powers in the matrices.
 For example, an appropriate choice is $n = - 4
m$ where we obtain a structure which is very close to the texture $T_4$ of Table 1. In
particular, choosing {$m=4,n=-16$}, the lower
$2\times 2$ charge entries in (\ref{qna}) are  positive,  while the rest are
negative so we have the following  structure of the up  and down quark mass
matrices:
\be
m_U=\left(\begin{array}{ccc}
\e^{16}&\e^6&\e^8\\
\e^6&\eb^4&\eb^2\\
\e^8&\eb^2&1\\
\end{array}\right);
\,\,\,
m_D=\left(\begin{array}{ccc}
\l^{16}&\l^6&\l^8\\
\l^6&\lb^4&\lb^2\\
\l^8&\lb^2&1\\
\end{array}\right)\,.
\label{upi3}
\ee

{\it $i_C$}:
We finally give the mass matrices for one more choice, which, as we will see
coincide with the one presented in ref~\cite{Ibanez:1994ig}. Under our definitions
of charges, this case arises if we put $m=2$ and $n=-8$,
\be
m_U=\left(\begin{array}{ccc}
\e^{8}&\e^3&\e^4\\
\e^3&\eb^2&\eb\\
\e^4&\eb&1\\
\end{array}\right);
\,\,\,
m_D=\left(\begin{array}{ccc}
\l^{8}&\l^3&\l^4\\
\l^3&\lb^2&\lb\\
\l^4&\lb&1\\
\end{array}\right)\,.
\label{upi4}
\ee
Notice however the appearance of two expansion parameters in~(\ref{upi4})
compared to only one used to appear in ref~\cite{Ibanez:1994ig}.

Above, we have provided examples based on a different charge assignment
($m,n$--values)  which naturally  give hierarchical patterns for the quark
sector. A natural question  now arises which of these cases fits better the
observed hierarchy and mixing effects.  There are mainly three sources of
further constraints that would definitely guide us to pick up
one definite case. First,
one needs an exact value of the parameter $\xi$ which  determines the singlet
higgs vevs. Second, the order one coefficients which are not  calculable, may
point to a certain choice. Finally, the structure of the lepton mass  matrix
will provide further information on the parameters $\l,\lb$. We proceed now
to  the other two possibilities for $m,n$.

{\bf case}{\it\, ii}: $ m$ odd, $n$ even.\\
This case assumes odd values for $m$ and even for $n$ which lead to an exact
texture--zero as in the case of Table 1. To obtain viable matrices, we may take
either $n= 2 m$ or $n= - 2 m$. These choices lead to   the same texture $T_3$
but with different expansion parameters. Taking
$m=3,n=\pm 6$    we get
\be
m_U=\left(\begin{array}{ccc}
\eb^{6}&0&\eb^3\\
0&\eb^3&0\\
\eb^3&0&1\\
\end{array}\right) \ {\rm and}\
m_U=\left(\begin{array}{ccc}
\e^6&0&\e^3\\
0&\eb^3&0\\
\e^3&0&1\\
\end{array}\right)\label{upii1}
\ee
respectively. How these zero entries arise? Bearing in mind that $m$ was taken to be
odd and $n$ even, the charge entries 11,12,21 and 23,32 in the charge--matrix
of  (\ref{qna}) are half--integers. Since the singlet charges are $\pm 1$, it
is not  possible to generate contributions in these entries from
non--renormalizable terms.  These are exact texture--zero mass matrices and
their form was proposed purely from  phenomenological analysis in
ref\cite{Giudice:1992an}. As in the first case discussed  above, the down
quark mass matrices of these two cases are obtained with the  replacements
$\e\ra \l$ and $\eb\ra \lb$, thus these are
\be
m_D=\left(\begin{array}{ccc}
\lb^{6}&0&\lb^3\\
0&\lb^3&0\\
\lb^3&0&1\\
\end{array}\right) \ {\rm and}\
m_D=\left(\begin{array}{ccc}
\l^6&0&\l^3\\
0&\lb^3&0\\
\l^3&0&1\\
\end{array}\right)\,.\label{doii1}
\ee
We will work out this case further, when we will discuss the corresponding lepton
matrix for total higgs charge  $h_+\ne 0$.

{\bf case} {\it \, iii:} $\,\, n,m$  odd.\\
We finally examine the case  where both $n,m$ are odd.  Here we obtain  mass
matrices similar to the up--quark texture $T_1$. We have the freedom to use
several sets of $m,n$ pairs. A suitable choice is  $m=-3,\, n=11$ which gives,
\be
m_U=\left(\begin{array}{ccc}
\eb^{11}&\eb^4&0\\
\eb^4&\e^3&0\\
0&0&1\\
\end{array}\right)\,.
\ee
A slightly different matrix involving only one parameter arises  for $n=5$
and $m=3$. It leads to the same texture zero, however different powers of
the expansion parameters appear. Ones gets,
\be
m_U=\left(\begin{array}{ccc}
\eb^{5}&\eb^4&0\\
\eb^4&\eb^3&0\\
0&0&1\\
\end{array}\right)\label{upiii1}
\ee

A general comment for the case $iii$ is necessary here:  due to the same
structure of the up and down quark mass matrices, the exact texture--zero mass
matrices in this case have small chance to reproduce the correct
Kobayashi--Maskawa (KM)--mixing.  Indeed, since the down quark mass matrix has
the same form  with the up, the KM--mixing of the third generation with the other
two --although  experimentally is measured to be small-- cannot be generated due
to the complete  decoupling of the third generation. However, in a realistic case,
as in string model  building, more than one pair of singlet fields acquire
non--zero vevs. Usually, some  other singlets with different charge assignment
form some higher order Yukawa  couplings with the fermions and generate small but
nevertheless important  contributions to the zero entries of the fermion
matrices. Another source of induced  small mixing arises from renormalization
group effects. If charged lepton and Dirac--neutrino Yukawa couplings are not
flavour diagonal (and as we will see, this is  exactly what happens in the
present case), then small calculable non--zero entries will  replace the zeros in
the above $m_U$ texture.

Having completed the analysis of the quark textures, we now need to consider
the implications on the lepton mass matrix structure. Closing this subsection
we simply note the remarkable fact that, even with one $U(1)_X$ anomalous
symmetry and only one pair of singlet fields one is able to reproduce four out
of the five phenomenological textures of Table 1.

{\bf \underline{Leptons}}

The analysis of the quark mass matrices  in the previous section,  has put
several constraints on the values of $m,n$--parameters. Note also that already
the phenomenological  constraint which implies  the successful relation
$m_{\tau}=m_b$ at the unification scale has also been imposed on the
$U(1)_X$ charges. Thus, the only remaining freedom to construct the charged
lepton mass matrices in the case $h_+=0$ is the value of the parameter $k$.
Note further, that there is no freedom to adjust the 12,21 elements of the
charge--lepton matrix since they are fixed completely by the quark matrix.
Bearing in mind that the parameters $m,n$ are integers, we can easily  see that
only the case of integer values $k=0,\pm1,\pm2,\dots$ can lead to acceptable
lepton mass matrices. In the following, we examine viable lepton textures with
respect to the  value of $k$ for each of the three cases in the quark sector
discussed above.

{\it\, i :} As in the corresponding case for quarks, we derive here the
lepton mass matrices for three $(m,n)$--sets and viable choices for $k$.\\
{\it $i_A$}: For $n = 2 m > 0$, we may take for example $m=4$, $n=8$ and $k=0$
or $k=-1$, so we obtain
\ba
m_L=\left(\begin{array}{ccc}
\lb^8&\lb^6&\lb^4\\
\lb^6&\lb^4&\lb^2\\
\lb^4&\lb^2&1\\
\end{array}\right),\,
{\rm or}&
m_L=\left(\begin{array}{ccc}
\lb^{7}&\lb^6&0\\
\lb^6&\lb^3&0\\
0&0&1\\
\end{array}\right)
\label{lt1}
\ea
respectively.
These correspond to the approximate hierarchies $m_e: m_\mu :
m_\tau\approx\lb^8:\lb^4:1$ or $m_e:m_\mu:m_\tau\approx\lb^7:\lb^3:1$.
The first matrix predicts exactly the  correct relation $\det M_D=\det
M_L$~\cite{Georgi:1979df} while the second gives also a quite satisfactory
result up to order one coefficients.

{\it $i_B$}:
 Let us now take $n=-4 m $, with the additional restriction that $m+k>0$.
Certainly, by inspection of the charge--matrix (\ref{lep}) we conclude that the
entries 22,23,32 have positive charges whilst all entries connected to the first
generation obtain negative ones. This means that to lowest order, we can
cancel the charge of the first with $\lb$--powers and the charge of the second
with powers of the expansion parameter $\l$. Thus, the following texture arises
\be
m_L=\left(\begin{array}{ccc}
\l^{(4m+k)}&\l^{3m/2}&\l^{(4m+k)/2}\\
\l^{3m/2}&\lb^{m+k}&\lb^{(m+k)/2}\\
\l^{(4m+k)/2}&\lb^{(m+k)/2}&1\\
\end{array}\right)
\label{lt2}
\ee
Choosing now $m=4$ (as in the corresponding quark case) and $k=-1$ we arrive to
a texture--zero  matrix of the form,
\be
m_L=\left(\begin{array}{ccc}
\l^{15}&\l^6&0\\
\l^6&\lb^3&0\\
0&0&1\\
\end{array}\right)
\label{lt2b}
\ee
Comparing with the previous case (\ref{lt2}) we see that we have now the possibility
of adjusting the value of the $22$--entry independently from the other matrix
elements. Indeed, recall that the down quark mass hierarchy in this case is
$\lambda^{12}:\bar\lambda^4 :1 $ which implies the hierarchical relation
$\bar\lambda \approx \lambda^{3/2}$ to $\lambda^2$ (depending on the order one
coefficients). This relation fits also well the charged--lepton mass hierarchy.\\
{\it
$i_C$}: Finally, we derive the lepton matrix which corresponds to the case $i_C$ of
the quark sector. For $m=2$, $n=-8$ and $k=3$ we obtain
\be
m_L=\left(\begin{array}{ccc}
\l^{5}&\l^3&0\\
\l^3&\l^1&0\\
0&0&1\\
\end{array}\right)
\label{lt2r}
\ee
in accordance with the texture derived in~\cite{Ibanez:1994ig}.
We note also that we may have more possibilities by choosing  $m+k<0$,
obtaining  a different lepton  structure but such cases will not be
elaborated here.

{\it ii}:
 Here, as in the case of quark mass matrices, we take the cases $m=3, n = 6$
(the case
$m=3, n = -6$ can be worked out similarly).
Now we are free to choose the value of $k$ in order  to obtain a
natural charged--lepton mass hierarchy. Assuming $k$--values in the range $-3<k<6$
we can write the lepton mass matrix in form
\be
m_L=\left(\begin{array}{ccc}
\lb^{6-k}&0&\lb^{\frac{6-k}2}\\
0&\l^{3+k}&\l^{\frac{1+k}2}\\
\lb^{\frac{6-k}2}&\l^{\frac{1+k}2}&1\\
\end{array}\right).
\label{lt3}
\ee
An  interesting texture arises for the $k=-2$. This gives a lepton matrix which
has the same  structure with the quarks:
\ba
m_L=\left(\begin{array}{ccc}
\lb^{8}&0&\lb^{4}\\
0&\lb&0\\
\lb^{4}&0&1\\
\end{array}\right).
\label{lt4b}
\ea
This matrix gives eigenvalues in the ratios $-\lb^8:\lb$ to be compared with
the mass eigenstates $m_e/m_{\tau}:m_{\mu}/m_{\tau} $ at the unification scale.
We note however, that this relation is satisfied for a rather large value
of the expansion parameter $\lb$. Further, for $k=-1$ we obtain
\ba
m_L=\left(\begin{array}{ccc}
\lb^{7}&0&0\\
0&\lb^2&\lb\\
0&\lb&1\\
\end{array}\right).
\label{lt4bb}
\ea
which gives the ratios $-\lb^7:\lb^2$ for  $m_e/m_{\tau}:m_{\mu}/m_{\tau} $.
We will see soon that the matrices obtained for the case $(ii)$ are
phenomenologically more promising when we assume $h_+\ne 0$.
We note here that this texture implies large mixing in the $\mu-\tau$ sector and
it  could be distinguished from the first one, due to the different  flavour
violating processes it implies. In particular, we should expect an enhancement of
the  $\tau\ra \mu\gamma$ branching ratio compared to the first case.

{\it iii}:
In this last case we choose $n=11, m=-3$. We have observed that in the quark sector
there is no mixing between the two heavier generations. In contrast, in the case of
charged leptons this mixing may arise from a suitable choice of the additional
parameter $k$. Thus, for the quark matrix (\ref{upiii1}) choosing $k=1$ we obtain
\ba
m_L=\left(\begin{array}{ccc}
\lb^{10}&\lb^4&\lb^{5}\\
\lb^4&\l^{2}&\l^{2}\\
\lb^{5}&\l^{2}&1\\
\end{array}\right)\,.
\label{lt5a}
\ea

\subsection{$h_+\ne 0$}

We come now to the most general case where the sum of the higgs doublet charges
is different than zero. As explained in  Section~\ref{secc}, there are two
solutions of the anomaly equations under the symmetry requirements  and the
tree--level constraints. The full $U(1)_X$--charge assignment  of the two
solutions for the matter and higgs fields are now shown in Tables \ref{tsol0} and
 \ref{tsol1}.

It is a welcoming fact that the quark mass matrices (as can be easily checked),
do not change at all under this generalization thus, our analysis concerning the
up and down quark textures remains intact. We therefore turn our attention to the
case of the charged lepton mass matrices. In this case we can easily see that
only the entries connected with the first generation in (\ref{lep}) receive
additional contribution. The charge--lepton matrix in the general case
$h_+\ne0$ becomes,
\begin{equation}
C^{LE^cH_1}=\left(\begin{array}{ccc} n-k-h_+&\frac{m+n-h_+}{2}&\frac{n-k-h_+}{2}\\
\frac{m+n-h_+}{2}&m+k&\frac{m+k}{2}\\
\frac{n-k-h_+}{2}&\frac{m+k}{2}&0\\
\end{array}\right)
\label{lep1}
\end{equation}
with the replacement now of a new value for $k=2\ell_2+6 q_3-m+\frac{7}{3}h_{+}$.

There is an additional contribution which equals the minus sum of the higgs
charge ($-h_+$) in the entries $11, 12,/21$ and $13, /31$  thus, in the general
case the elements 12 and 21 are no--longer equal to the corresponding ones of the
quark matrix. Our notation here might be confusing in the sense that there appear
four different parameters in the  lepton case, namely $m,n,k$ and $h_+$. In fact,
(as it is clear from (\ref{lcharge})), there are only two parameters
$\ell_1-\ell_3$ and $\ell_2-\ell_3$  which enter in this structure;
here they can be taken to be the combinations $n+m-h_+$ and $m+k$.
This is the price we have to pay in order to keep the  parametrization already
used,  and transfer the constraints from the quark sector.

In the above parametrization we can easily see now that the case $h_+\ne 0$  has
some important implications on the lepton mass matrix structure. First, from our
analysis in the quark sector we observe that we are forced to take integer values
for  the parameters $m,n$. We can  easily see that a non--integer value of the
total higgs charge $h_+$ would lead to a  massless state. As a result we are
forced to assume only integer values for  both, $k$ and $h_+$ parameters.

Let us consider now the first Solution (\ref{sol0}). As seen from Table
\ref{tsol0} the sum of the higgs doublet charges is fixed $h_+=m+n$ and thus the
elements 12 and 21 of the lepton matrix (\ref{lep1}) vanish. This means that the
associated couplings become of the order of the $\tau$ mass and this texture
leads to  two heavy eigenstates.  Thus we will not consider this solution further
and from this point we will refer only to Solution (\ref{h2eq}) when discussing
$h_+\ne0$.

Another important constraint arises from the relation $\det m_D=\det m_L$
\cite{Georgi:1979df}. Assuming~\footnote{Similar relations are obtained for
$n<0$ by interchanging $\e\leftrightarrow\eb$.} $n>0$ we have
\ba
\det m_D=\left\{
\begin{array}{l}
\l^{-m} \lb^{n} , m<0, m+n={\rm odd}\\
\lb^{m+n}\ or\ \l^{- m-n}\ {\rm otherwise}
\end{array}
\right.
\ea
for the quarks. Similarly for $k+n-h_+>0$ we have
\ba
\det m_L=\left\{
\begin{array}{l}
\l^{-k-m}\, \lb^{n-k-h_+} , k+m<0, m+n-h_+={\rm odd}\\
\lb^{m+n-h_+}\ {\rm or}\ \l^{-m-n+h_+} ,\ {\rm otherwise}
\end{array}
\right.
\ea
The eigenvalues of the lepton mass matrix can be also worked out. They have the form
\ba
(\l\,\, {\rm or}\,\, \lb)^{|h_+ +k -n|}\,,(\l\,\, {\rm or}\,\, \lb)^{|k+m|},\,\, 1
\ea
Notice that the presence of $h_+$ affects only the lightest eigenvalue.

We wish now to give one more example where we can obtain a realistic texture--zero
matrix. We choose the values $m=1,n=2$ which correspond to  quark matrices
 of case type $T_3$  as in $ii$. Taking $k=-2$ and $h_+=9$ we obtain
\ba
m_D=\left(\begin{array}{ccc}
\lb^{2}&0&\lb\\
0&\lb&0\\
\lb&0&1\\
\end{array}\right),
&
m_L=\left(\begin{array}{ccc}
\l^{5}&\l^3&0\\
\l^3&\l&0\\
0&0&1\\
\end{array}\right)
\label{lt6}
\ea
while  $m_U$ has the same texture--zero as $m_D$ provided that we replace $\l\ra
\e$. The above texture--zero charged lepton matrix is different from the
$m_D$--matrix. It implies no mixing for the $\tau$ lepton while it predicts the
correct hierarchy, provided we impose the relation $\lb\approx \l^2$.

We give a final example by taking $m=3, n= 6$, $h_+=12$, $h_1=1$, $k=-2$ and
$q_3=-4$. Then, we obtain the same texture--zero for both, down quark and lepton
matrices:
\ba
m_D=\left(\begin{array}{ccc}
\lb^{6}&0&\lb^3\\
0&\lb^3&0\\
\lb^3&0&1\\
\end{array}\right),
&
m_L=\left(\begin{array}{ccc}
\l^{4}&0&\l^2\\
0&\lb&0\\
\l^2&0&1\\
\end{array}\right).
\label{lt6A}
\ea

\section{Baryon and Lepton number violating operators.}

In addition to the standard Yukawa couplings which provide with masses quarks and
leptons, the gauge symmetry of the MSSM allows also terms which violate baryon
and lepton number already at the tree--level. Suppressing generation indices, the
terms relevant to proton decay are written
\ba
\l L L E^c + \l' L Q D^c + \l'' U^c D^c D^c
\label{blv}
\ea
There are also gauge invariant higgs, and lepton--higgs mixing
terms of the form
\ba
\mu H_1 H_2 + \mu' L H_2
\label{hmx}
\ea
 If  all terms (\ref{blv})  are allowed in the superpotential they lead to
fast proton decay. In particular, the combination of the terms $ L Q D^c$
and $ U^c D^c D^c$  generates an effective dimension four operator via the
diagram generated by exchanging the scalar component of the $D^c$ superfield.
Imposing the $R$--parity~\cite{Farrar:1978xj,Farrar:1983te} multiplicative
symmetry  $R = (-1)^{3 B + 2 S + L}$ under which matter fields (quarks and
leptons)  change sign while the higgs doublets transform to themselves, all
dangerous  terms change sign and are eliminated from the superpotential.

 $R$--parity prevents also the appearance of the second higgs mixing term
 in (\ref{hmx}). However, the usual $\mu$--term, i.e., the direct mixing
 between the two electroweak higgs fields is invariant under the
 $R$--symmetry. In the model under consideration this may lead to a disaster,
 as this mixing can be generated by a term of the form
 $\phi^s\bar\phi^r H_1 H_2$ where $r,s$ are suitable powers  matching
 the sum of the charge of the two higgs doublets. With vevs $\langle\phi\rangle,
\langle\bar\phi\rangle\sim 10^{-1} M_U$
 --as required by the $D$--term cancellation condition and the fermion
 mass textures-- a large power (at least $r+s> 15$)
 is needed to suppress sufficiently
 the $\mu$--mass parameter  and bring it down to the electroweak scale.

In addition to the tree--level couplings there are also higher
order gauge invariant terms leading to  dangerous dimension--five
operators which  induce proton decay. The ones surviving
$R-$parity  are
\cite{IR}
\ba
\frac{\lambda_4^{ijkl}}{M_U}\,Q_iQ_jQ_kL_l,&
\displaystyle\frac{\lambda_5^{ijkl}}{M_U}\,U^c_iU^c_jD^c_k E^c_l\label{dim5}
\ea
where the indices $i,j,k,l =1,2,3$ refer to the three generations.
Although the induced amplitudes of dimension--five operators are
relatively suppressed compared to those arising from the
(\ref{blv}) terms, due to the fact that they arise as
non--renormalizable interactions, the baryon decay bounds on their
Yukawa coupling constants are very restrictive. In the general
case one has to impose $\lambda_4 <10^{-7}$ for operators
involving light quarks while the constraints are less important
for $\lambda_5$
\cite{IR}. If an expansion parameter $\eps \sim 0.23$ is involved in the coupling,
we should require a power $\eps^{9}$ for a coupling involving only
first and second  generation fermions to comply with the
experimental bound. Couplings involving third generation fields
suffer additional suppression from mixing angles and the bounds
are less restrictive. Therefore it is crucial to examine whether
the charge assignment of the fermion fields under the anomalous
$U(1)_X$ symmetry is also capable of eliminating these baryon
and lepton number violating operators.

We have classified all possible non--zero couplings involving the
various generations together with their $U(1)_X$ charges and
exhibit them  in Table \ref{tb5}. The total $U(1)_X$ charge of
each operator is now expressed only in terms of the free
parameters $m,n,k$ and the sum of the higgs charges $h_+=h_1+h_2$.

In the first column of this table we write the particular operator in terms of its
family indices while in the second column we present its charge. Since everything here
is parametrized in terms of the charge of the singlet, we should simply check whether
the charge of a particular operator is integer or non--integer. We now distinguish
two cases:

\begin{itemize}

\item

{\bf \underline{A: $h_1+h_2\equiv h_+\equiv 0$}}

To analyse the effects of the anomalous abelian symmetry on these operators, let us
start with the case $h_+= h_1+h_2=0$. As seen in Table \ref{tb5} the charges of these
operators depend only on the integer parameters $m,n,k$ and they involve $1/2$
fractions of these parameters. Therefore, we consider which of these operators survive
for various choices of $m,n,k$. We assume that the value of the parameter $k$ is odd.
This choice of $k$ fits perfectly with the findings in the lepton mass matrices.
(Indeed, in Section 3, most of the acceptable lepton mass textures where constructed
choosing odd values for $k$.) Clearly, the most favorable case is when both $m$ and
$n$ are even as it eliminates most of the operators involving the light generations.
The rest of the operators are needed to be suppressed with appropriate selection of
$m$ and $n$ and $k$. As it will become clear in the next section one can easily find
charge assignments (see e.g. Solution A of Table~\ref{sols}) that give acceptable
fermion mass textures and adequately suppress all these operators.

\item
{\bf \underline{B: $h_1+h_2\equiv h_+\ne 0$}}

Now, let us come to the most general case. It is interesting that the
dimension--5 proton decay operators can also be expressed in terms of integer
parameters, namely $m,n,k,h_+$ and they do not involve $h_2$.  Actually the
dimension--5 operators of Table~\ref{tb5} are receiving additional charge, the
first seven (of the form $QQQL$) obtain an $-\frac{5}{3}h_{+}$ additional charge
while the remaining  receive a contribution of $-\frac{7}{6}h_{+}$. The charges
of the operators of the type $U^cU^cD^cE^c$ are obtained by adding
$\frac{h_{+}}{3}$ to the charge of the  $QQQL$ operator in the same line.

We can choose the higgs charges so that the contributions $\frac 53 h_+$ and
$\frac 76 h_+$ are neither integers nor half--integers. Then, all operators are
eliminated simultaneously.
\end{itemize}

Another non--renormalizable operator allowed by $R$--parity is the 
following~\cite{Barbieri:1980hc}
\ba
\frac{\lambda_8}{M_U} (L_i\bar H)(L_j\bar H)\label{maj}
\ea
where $i,j$ refer to generations. This operator which violates lepton number by
two units, may have interesting phenomenological consequences as it is capable of
generating a Majorana mass for the left--handed neutrino. A coupling
$\lambda_8\approx 1- 10^{-2}$ would be of the right order for such a mass term.
In Table~\ref{tlhlh} we present all relative operators and their charges for the
case $h_+=0$. When $h_+\ne0$ the relative charges can be calculated using Table
\ref{tsol1}. The role of this term in specific examples will be
presented in the next section.

A more difficult problem however is related to the $\mu$--term. As is well
known, there must be a higgs mixing via a term of the form $\mu h\bar h$ with
$\mu\sim m_W$ in order to prevent the appearance of an unwanted  axion. In the
simple scenario of one $U(1)$ symmetry and the two singlet fields we discuss
here, this
is not easy. In general, if the charge $h_+=h_1+h_2$ is an integer, then the
singlet fields $\phi, \bar\phi$ may couple to the combination $h\bar h$, giving
rise to a $\mu$ `mass'--parameter of the
order $\left(\bar\phi/M_U\right)^{h_+-1}\phi$!! Since the vev of
$\langle\phi\rangle \sim 10^{-1} M_U$  one has to impose the condition
$h_+\ge 15 $, otherwise the higgs doublets receive unacceptably large
masses\cite{Ellis:1999ce}.

There are also other possible ways of avoiding such a large mass term for the
higgs doublets. For example, one may introduce a Peccei--Quinn
symmetry~\cite{PQ}
 to ban~\cite{Ibanez:1999it} simultaneously the higgs mixing as well as
the proton decay operators discussed above. In our case, since the higgs charges
are basically unconstrained, it is  possible to work out cases where  their sum is
not an integer. Therefore, the higgs term does not appear. We note however, that
solutions which eliminate completely the $\mu$--term are not favourable; if a term
is completely forbidden for symmetry reasons in the superpotential, it is not
obvious how it can appear in the K\"alher potential. We think that the
suppression of the higgs mixing coupling by an appropriate choice of the higgs
charges is a rather natural solution. In Table (\ref{sols}) we give cases
with field-- charges which lead to a large $\mu$--term suppression and a viable
set of Yukawa mass matrices.

 We note that, even if we
ignore the above problem of the higgs mixing --assuming the existence of another
type of solution-- and impose a half--integer value of $h_+$, we encounter another
difficulty; we know from the analysis of the quark mass matrices that $m,n$ are
integers while from the lower $2\times 2$
charged--lepton mass matrix, we find that $k$ also
has to be integer. Then, we infer that  the non--integer values of $h_+$ lead
unavoidably to a massless electron state. In a more complicated theory we may
hope that radiative effects or  other weakly--coupled singlets could
generate a small entry adequate to provide the electron with a mass.

We would like now to abandon the $R$--parity symmetry and investigate the possibility
of constructing a set of charges which give viable fermion mass textures with baryon
and lepton violation within the existing limits.
In Tables \ref{tlqd},\ref{tlle}, \ref{tudd} we present all dangerous trilinear
operators capable of inducing proton decay. In the second column we exhibit their
total charge under the
$U(1)_X$ anomalous symmetry. We have expressed the total charge in terms of the
parameters
$m,n,k$ (which parameterize all quark and charged lepton mass matrices) and the charge
of the third generation quark doublet $q_3$. Thus, in order to generate a gauge
invariant baryon violating term we should be able to add a singlet $\phi$ or
antisinglet $\bar\phi$ to the proper power $r$, $\phi^r$ ($\bar\phi^r$) to cancel the
charge. E.g, if
$q_i+q_j+\ell_k=\pm r$, then the operator $Q_iQ_jL_k\bar\phi^r$ ($\phi^r$)
cannot be avoided.


\begin{table}
\centering
\begin{tabular}{|l|c|c|c|}
\hline
field&\multicolumn{3}{c|}{generation}\\
\hline
&1&2&3\\
\hline
$Q$&$\frac{n}{2}+q_3$&$\frac{m}{2}+q_3$&$q_3$\\
$U^c$&$\frac{n}{2}+q_3$&$\frac{m}{2}+q_3$&$q_3$\\
$D^c$&$\frac{n}{2}-3q_3$&$\frac{m}{2}-3q_3$&$-3q_3$\\
$L$&$\frac{n-k}{2}-3q_3$&$\frac{m+k}{2}-3q_3$&$-3q_3$\\
$E^c$&$\frac{n-k}{2}+q_3$&$\frac{m+k}{2}+q_3$&$q_3$\\
\hline
\multicolumn{4}{|c|}{Higgs}\\
\hline
$H_1$&$2q_3$&$H_2$&$-2q_3$\\
\hline
\end{tabular}
\caption{$U(1)_X$ charge assignments for the case $h_{+}=0$}
\label{thp0}
\end{table}
\begin{table}
\centering
\begin{tabular}{|l|c|}
\hline
Operator& ${U(1)}_X$ charge\\
\hline
$L_1 H_2 L_1 H_2$&$n-k-10q_3$\\
\hline
$L_1 H_2 L_2 H_2$&$\frac{m+n}{2}-10q_3$\\
\hline
$L_1 H_2 L_3 H_2$&$\frac{n-k}{2}-10q_3$\\
\hline
$L_2 H_2 L_2 H_2$&$m+k-10q_3$\\
\hline
$L_1 H_2 L_3 H_2$&$\frac{k+m}{2}-10q_3$\\
\hline
$L_3 H_2 L_3 H_2$&$-10q_3$\\
\hline
\end{tabular}
\caption{\label{tlhlh}$U(1)_X$ charges of the operators $L_i L_j H_2 H_2$
for the case $h_{+}=0$. Indices refer to generations.}
\end{table}
 In this case, the Yukawa couplings of
the terms (\ref{blv}) should be highly suppressed, in particular those involving first
generation quark and lepton states. According to our natural assumption that the
non--calculable coefficients should be of order one, we infer that the $U(1)_X$
symmetry should prevent the appearance of such terms at the renormalizable
superpotential. These operators should appear at high orders so that their
couplings are suppressed by proper powers of the expansion parameter. In order to
put appropriate constraints on the $U(1)_X$--charges, we first need the
experimental bounds on the relevant Yukawa couplings. The most severe bounds are
imposed on the Yukawa couplings $\lambda_{111}'$ and $\lambda_{133}'$ of this
operator. In particular, from the absence of the exotic reaction of
$\beta\beta$--decay we have $\lambda_{111}'< 10^{-3}$ and
from the bounds on the left--handed neutrino Majorana mass, 
 $\lambda_{133}'< 2\times 10^{-3}$.  Other exotic decays
imply bounds to various combinations of couplings, while more restrictive
bounds arise for products of the form $\lambda\lambda'$;
 a recent analysis on  the various Yukawa
couplings predicted in $U(1)$ models  and a relevant discussion of the above operators
can be found  in~\cite{Ellis:1998rj}.

Note also that, when $R$--parity is absent, additional dimension--five operators
involving higgs multiplets are also dangerous when they are combined with the
couplings (\ref{blv}) leading to proton decay via loop-graphs (For a complete list of
these operators see \cite{IR}). In particular, operators of the form $[QQQH_1]_F$ are
dangerous in the presence of
$LQD^c$ couplings while the operators $[QU^cE^cH_1]_F$ are also dangerous in the
presence of $U^cD^cD^c$ terms. The former, leads to a tree--level proton decay diagram
via the higgs vev $H_1$ and its coupling to the down quark $Q D^c H_1$, and
similarly the second leads to an effective $U^cU^cD^cE^c$ operator. Finally, one
should avoid the simultaneous existence of the term $U^cD^cD^c$ with the lepton number
violating operators
$[QU^cL^*]_D$ and $[QU^cL^*]_D$. It is now straightforward to turn the above bounds to
constraints on the $U(1)_X$ charges. Since in our subsequent analysis we will present
cases where all the tree--level operators are either suppressed, or banned by the
symmetry, we will not pursue this issue further.

\begin{table}
\centering
\begin{tabular}{|l|c|}
\hline
Operator& ${U(1)}_X$ charge\\
\hline
$L_1Q_1D^c_1$&${\frac{ 3\,n-k}{2}}- 5\,q_3$\\
$L_1Q_1D^c_2$, $L_1Q_2D^c_1$&$ n + {\frac{m-k}{2}}- 5\,q_3$\\
$L_1Q_1D^c_3$, $L_1Q_3D^c_1$&$ n - {\frac{k}{2}} - 5\,q_3$\\
$L_1Q_2D^c_2$&$ m + {\frac{n-k}{2}} - 5\,q_3$\\
$L_1Q_2D^c_3$, $L_1Q_3D^c_2$&${\frac{ m + n - k}{2}}- 5\,q_3$\\
$L_1Q_3D^c_3$&${\frac{n-k}{2}}- 5\,q_3$\\
$L_2Q_1D^c_1$&$n + {\frac{k + m }{2}}- 5\,q_3$\\
$L_2Q_1D^c_2$,$L_2Q_2D^c_1$&$m+{\frac{k +  n }{2}}- 5\,q_3$\\
$L_2Q_1D^c_3$, $L_2Q_3D^c_1$&${\frac{k + m + n }{2}}- 5\,q_3$\\
$L_2Q_2D^c_2$&${\frac{k + 3\,m }{2}}- 5\,q_3$\\
$L_2Q_2D^c_3$, $L_2Q_3D^c_2$&$m+{\frac{k}{2}} - 5\,q_3$\\
$L_2Q_3D^c_3$&${\frac{k + m }{2}}- 5\,q_3$\\
$L_3Q_1D^c_1$&$n - 5\,q_3$\\
$L_3Q_1D^c_2$, $L_3Q_2D^c_1$&${\frac{m + n}{2}}- 5\,q_3$\\
$L_3Q_1D^c_3$, $L_3Q_3D^c_1$&${\frac{n}{2}} - 5\,q_3$\\
$L_3Q_2D^c_2$&$m - 5\,q_3$\\
$L_3Q_2D^c_3$, $L_3Q_3D^c_2$&${\frac{m}{2}} - 5\,q_3$\\
$L_3Q_3D^c_3$&$-5\,q_3$\\
\hline
\end{tabular}
\caption{\label{tlqd}${U(1)}_X$  charges of the R-parity violating couplings
 $L_i Q_j D^c_k$ for the case $h_+=0$. The indices refer to the generations.}
\end{table}
\begin{table}
\centering
\begin{tabular}{|c|c|}
\hline
Operator& ${U(1)}_X$ charge\\
\hline
$L_1L_2E^c_1$&$n+{\frac{m-k}{2}} - 5\,q_3$\\
$L_1L_2E^c_2$&$m+{\frac{n +  k }{2}}-5\,q_3$\\
$L_1L_2E^c_3$&${\frac{m + n}{2}}- 5\,q_3$\\
$L_1L_3E^c_1$&$n-k  - 5\,q_3$\\
$L_1L_3E^c_2$&${\frac{m + n }{2}}- 5\,q_3$\\
$L_1L_3E^c_3$&${\frac{n-k}{2}}- 5\,q_3$\\
$L_2L_3E^c_1$&${\frac{m + n}{2}}-5\,q_3$\\
$L_2L_3E^c_2$&$k + m - 5\,q_3$\\
$L_2L_3E^c_3$&${\frac{k + m }{2}}- 5\,q_3$\\
\hline
\end{tabular}
\caption{\label{tlle}${U(1)}_X$  charges of the R-parity violating couplings
 $L_i L_j E^c_k$ for the case $h_+=0$. The indices refer to the generations.}
\end{table}
\begin{table}
\centering
\begin{tabular}{|c|c|}
\hline
Operator& ${U(1)}_X$ charge\\
\hline
$U^c_1D^c_1D^c_2$&$n + {\frac{m}{2}} - 5\,q_3$\\
$U^c_1D^c_1D^c_3$&$n - 5\,q_3$\\
$U^c_1D^c_2D^c_3$&${\frac{m + n}{2}}-5\,q_3$\\
$U^c_2D^c_1D^c_2$&$m + {\frac{n}{2}} - 5\,q_3$\\
$U^c_2D^c_1D^c_3$&${\frac{m + n }{2}}- 5\,q_3$\\
$U^c_2D^c_2D^c_3$&$m - 5\,q_3$\\
$U^c_3D^c_1D^c_2$&${\frac{m + n}{2}}- 5\,q_3$\\
$U^c_3D^c_1D^c_3$&${\frac{n}{2}} - 5\,q_3$\\
$U^c_3D^c_2D^c_3$&${\frac{m}{2}} - 5\,q_3$\\
\hline
\end{tabular}
\caption{\label{tudd}${U(1)}_X$  charges of the $R$--parity violating couplings
 $U^c_i D^c_j D^c_k$ for the case $h_+=0$. The indices refer to the generations.}
\end{table}

\begin{table}
\centering
\begin{tabular}{|l|l|c|c|c|c|}
\hline
\multicolumn{2}{|c|}{Operator}&$U(1)_X$--Charge & $m,n$ even& $m$ odd, $n$ even&
$n,m$ odd\\
\hline
$Q_1 Q_1 Q_2 L_1$ &$D^c_1 U^c_1 U^c_2 E_1^c$& $\frac{3n+m-k}{2}$ &&$\surd$   &  \\
\hline
$Q_1 Q_1 Q_3 L_1$ &$D^c_1 U^c_1 U^c_3 E_1^c$& $\frac{3n-k}{2}$ &&  &$\surd$  \\
\hline
$Q_1 Q_2 Q_3 L_1$ &
\begin{minipage}{2cm}
$D^c_1 U^c_2 U^c_3 E_1^c$\\
$D^c_2 U^c_1 U^c_3 E_1^c$\\
$D^c_3 U^c_1 U^c_2 E_1^c$
\end{minipage}
& $n+\frac{m-k}{2}$ & &$\surd$&$\surd$ \\
\hline
$Q_2 Q_1 Q_2 L_1$ &$D^c_2 U^c_1 U^c_2 E_1^c$& $m+n-\frac k2 $ && &     \\
\hline
$Q_2 Q_2 Q_3 L_1$ &$D^c_2 U^c_2 U^c_3 E_1^c$& $m+\frac{n-k}{2} $ && &$\surd$ \\
\hline
$Q_3 Q_1 Q_3 L_1$ &$D^c_3 U^c_1 U^c_3 E_1^c$& $n-\frac k2 $ &&   &     \\
\hline
$Q_3 Q_2 Q_3 L_1$ &$D^c_3 U^c_2 U^c_3 E_1^c$& $\frac{m+n-k}{2}$ &&$\surd$  & \\
\hline
$Q_1 Q_1 Q_2 L_2$ & $D^c_1 U^c_1 U^c_2 E_2^c$& $n+m+\frac k2$&& &\\
\hline
$Q_1 Q_1 Q_3 L_2$ & $D^c_1 U^c_1 U^c_3 E_2^c$& $n+\frac{k+m}{2}$&&$\surd$&$\surd$\\
\hline
$Q_1 Q_2 Q_3 L_2$ &
\begin{minipage}{2cm}
$D^c_1 U^c_2 U^c_3 E_2^c$\\
$D^c_2 U^c_1 U^c_3 E_2^c$\\
$D^c_3 U^c_1 U^c_2 E_2^c$
\end{minipage}
& $m+\frac{k+n}{2}$ & &&$\surd$ \\
\hline
$Q_2 Q_1 Q_2 L_2$ &$D^c_2 U^c_1 U^c_2 E_2^c$& $ \frac{n+3m+k}{2} $&&$\surd$ &\\
\hline
$Q_2 Q_2 Q_3 L_2$ &$D^c_2 U^c_2 U^c_3 E_2^c$& $ \frac{3m+k}{2} $& &$\surd$&$\surd$\\
\hline
$Q_3 Q_1 Q_3 L_2$ &$D^c_3 U^c_1 U^c_3 E_2^c$& $ \frac{m+k+n}{2} $& &$\surd$&\\
\hline
$Q_3 Q_2 Q_3 L_2$ &$D^c_3 U^c_2 U^c_3 E_2^c$& $ m+\frac k2 $& &&\\
\hline
$Q_1 Q_1 Q_2 L_3$ &$D^c_1 U^c_1 U^c_2 E_3^c$& $ n+\frac m2 $&$\surd$ &&\\
\hline
$Q_1 Q_1 Q_3 L_3$ &$D^c_1 U^c_1 U^c_3 E_3^c$& $n$&$\surd$&$\surd$ &$\surd$\\
\hline
$Q_1 Q_2 Q_3 L_3$ &
\begin{minipage}{2cm}
$D^c_1 U^c_2 U^c_3 E_3^c$\\
$D^c_2 U^c_1 U^c_3 E_3^c$\\
$D^c_3 U^c_1 U^c_2 E_3^c$
\end{minipage}
& $\frac{m+n}{2}$&$\surd$& &$\surd$\\
\hline
$Q_2 Q_1 Q_2 L_3$ &$D^c_2 U^c_1 U^c_2 E_3^c$& $ m+\frac n2 $&$\surd$ &$\surd$&\\
\hline
$Q_2 Q_2 Q_3 L_3$ &$D^c_2 U^c_2 U^c_3 E_3^c$& $m$&$\surd$&$\surd$ &$\surd$\\
\hline
$Q_3 Q_1 Q_3 L_3$ &$D^c_3 U^c_1 U^c_3 E_3^c$& $\frac{n}2 $&$\surd$& $\surd$&\\
\hline
$Q_3 Q_2 Q_3 L_3$ &$D^c_3 U^c_2 U^c_3 E_3^c$& $\frac{m}2 $&$\surd$& &\\
\hline
\end{tabular}
\caption{\label{tb5}
{ Dimension--five operators leading to proton decay are presented in
the first and second column. The associated charges for the case
$h_{+}=0$ are presented in the third column. The symbol $\surd$
marks the surviving operators for the allowed values of $m$ and
$n$ assuming $k=$odd. When $h_+\ne 0$, all $QQQL$ operators receive an
additional charge ($\frac 53 h_+$ or $\frac 76 h_+$), thus for
appropriate $h_+$ values are forbidden. The $D^cU^cU^cE^c$ receive
similar contributions. (For details see Section
6). } }
\end{table}

\section{ A few typical solutions}
\begin{table}
\centering
\begin{tabular}{|l|c|c|c|}
\hline
\multicolumn{4}{|c|}{Solution A}\\
\hline
field&\multicolumn{3}{c|}{generation}\\
\hline
&1&2&3\\
\hline
$Q$&$4$&$2$&$0$\\
$D^c$&$4$&$2$&$0$\\
$U^c$&$4$&$2$&$0$\\
$L$&$4$&$2$&$0$\\
$E^c$&$4$&$2$&$0$\\
\hline
\multicolumn{4}{|c|}{Higgs}\\
\hline
$H_1$&$0$&$H_2$&$0$\\
\hline
\multicolumn{4}{|c|}{Singlets}\\
\hline
$\phi$&$1$&$\bar\phi$&$-1$\\
\hline
\end{tabular}
\begin{tabular}{|l|c|c|c|}
\hline
\multicolumn{4}{|c|}{Solution B}\\
\hline
field&\multicolumn{3}{c|}{generation}\\
\hline
&1&2&3\\
\hline
$Q$&$\hphantom{+}6$&$\hphantom{+}4$&$\hphantom{+}2$\\
$D^c$&$-2$&$-4$&$-6$\\
$U^c$&$\hphantom{+}6$&$\hphantom{+}4$&$\hphantom{+}2$\\
$L$&$-2$&$-4$&$-6$\\
$E^c$&$\hphantom{+}6$&$\hphantom{+}4$&$\hphantom{+}2$\\
\hline
\multicolumn{4}{|c|}{Higgs}\\
\hline
$H_1$&$4$&$H_2$&$-4$\\
\hline
\multicolumn{4}{|c|}{Singlets}\\
\hline
$\phi$&$1$&$\bar\phi$&$-1$\\
\hline
\end{tabular}
\begin{tabular}{|l|c|c|c|}
\hline
\multicolumn{4}{|c|}{Solution C}\\
\hline
field&\multicolumn{3}{c|}{generation}\\
\hline
&1&2&3\\
\hline
$Q$&$\hphantom{+}\frac 92$&$\hphantom{+}\frac
52$&$\hphantom{+}\frac 12$\\ $D^c$&$\hphantom{-}\frac 52
$&$\hphantom{+}\frac 12$&${-}\frac 32$\\ $U^c$&$\hphantom{+}\frac
92$&$\hphantom{+}\frac 52$&$\hphantom{+}\frac 12$\\
$L$&$\hphantom{-}\frac 52$&$\hphantom{+}\frac 12$&${-}\frac 32$\\
$E^c$&$\hphantom{+}\frac 92$&$\hphantom{+}\frac
52$&$\hphantom{+}\frac 12$\\\hline
\multicolumn{4}{|c|}{Higgs}\\
\hline
$H_1$&$\hphantom{-}1$&$H_2$&$-1$\\
\hline
\multicolumn{4}{|c|}{Singlets}\\
\hline
$\phi$&$1$&$\bar\phi$&$-1$\\
\hline
\end{tabular}
\begin{tabular}{|l|c|c|c|}
\hline
\multicolumn{4}{|c|}{Solution D}\\
\hline
field&\multicolumn{3}{c|}{generation}\\
\hline
&1&2&3\\
\hline
$Q$&$\hphantom{+}\frac 73$&$\hphantom{+}\frac 56$&$-\frac 23$\\
$D^c$&$\hphantom{-}5$&$\hphantom{+}\frac 72$&$\hphantom{-}2$\\
$U^c$&$\hphantom{+}\frac 73$&$\hphantom{+}\frac 56$&$-\frac 23$\\
$L$&$\hphantom{-}5$&$\hphantom{+}\frac 72$&$\hphantom{-}2$\\
$E^c$&$\hphantom{+}\frac 73$&$\hphantom{+}\frac 56$&$-\frac 23$\\
\hline
\multicolumn{4}{|c|}{Higgs}\\
\hline
$H_1$&$-\frac{4}{3}$&$H_2$&$\hphantom{-}\frac 43$\\
\hline
\multicolumn{4}{|c|}{Singlets}\\
\hline
$\phi$&$1$&$\bar\phi$&$-1$\\
\hline
\end{tabular}
\begin{tabular}{|l|c|c|c|}
\hline
\multicolumn{4}{|c|}{Solution E}\\
\hline
field&\multicolumn{3}{c|}{generation}\\
\hline
&1&2&3\\
\hline
$Q$&$-\frac 12$&$- \frac 52$&$-\frac 92$\\
$D^c$&$-\frac{17}{6}$&$-\frac{29}{6}$&$-\frac{41}{6}$\\
$U^c$&$\hphantom{+}\frac {23}{6}$&$\hphantom{+}\frac {11}{6}$&$-\frac{1}{6}$\\
$L$&$-\frac{55}{6}$&$-\frac{19}{6}$&$-\frac {31}{6}$\\
$E^c$&$-\frac{61}{6}$&$-\frac{25}{6}$&$-\frac{37}{6}$\\
\hline
\multicolumn{4}{|c|}{Higgs}\\
\hline
$H_1$&$\hphantom{+}\frac {34}{3}$&$H_2$&$\hphantom{+}\frac {14}{3}$\\
\hline
\multicolumn{4}{|c|}{Singlets}\\
\hline
$\phi$&$1$&$\bar\phi$&$-1$\\
\hline
\end{tabular}
\caption{Some typical $U(1)_X$ charge assignments consistent with anomaly
cancellation and acceptable fermion mass matrices.\label{sols}}
\end{table}

We now pass to an investigation of possible solutions which are in accordance
with the phenomenological requirements discussed in the previous section.
There are numerous case of $U(1)_X$ charge assignments which give textures consistent
with the hierarchical fermion mass pattern. Here, we present only few 
charachteristic  examples  which mainly  fall into
two categories: Those, which allow baryon  and lepton number
violating operators and need
additional underlying symmetries to evade them and, those which strictly
forbid any lepton and baryon violating operator.
\begin{itemize}
\item

{\bf Solution A}\\ It is a remarkable fact that one of the most promising
texture--zero mass matrices found in Section 3 arises from a simple generation
independent charge assignment. The first generation fermions are assigned with charge
4, the second with 2 and third with 0 (see table \ref{sols}). This yields (\ref{upi1})
for the up quarks
\be
m_U=\left(\begin{array}{ccc}
\eb^8&\eb^6&\eb^4\\
\eb^6&\eb^4&\eb^2\\
\eb^4&\eb^2&1\\
\end{array}\right),
\ee
and similarly for down quarks and leptons
\be
m_L \sim m_D = \left(\begin{array}{ccc}
\lb^8&\lb^6&\lb^4\\
\lb^6&\lb^4&\lb^2\\
\lb^4&\lb^2&1\\
\end{array}\right)
\ee
The above charge assignment although it allows  dimension--5
operators, it  sufficiently suppresses the dangerous ones. The
suppression factors are
\ba
\lambda_4^{3233}\sim \lb^2\nonumber\\
\lambda_4^{3133}, \lambda_4^{3223}, \lambda_4^{3232}\sim \lb^4\nonumber\\
\lambda_4^{3231}, \lambda_4^{3222}, \lambda_4^{3132}, \lambda_4^{1233}
\sim \lb^6\nonumber\\
\lambda_4^{2231}, \lambda_4^{3131}, \lambda_4^{1232},
\lambda_4^{1133},
\lambda_4^{2123}
\sim \lb^8\label{sup5}\\
\lambda_4^{1231}, \lambda_4^{1132}, \lambda_4^{2122}, \lambda_4^{1123}\sim \lb^{10}\nonumber\\
\lambda_4^{1131},\lambda_4^{2121},\lambda_4^{1122}\sim \lb^{12}\nonumber\\
\lambda_4^{1121}\sim \lb^{14}\nonumber
\ea
and similarly for $\lambda_5^{ijkl}$, where the couplings refer to
Eq. (\ref{dim5}).

This solution has also the advantage of {\it not} suppressing the
quartic couplings $L_iL_jH_2H_2$. Actually they have the form
\be
L_i L_j H_2 H_2\sim \frac{m_W^2}{M_U} \left(\begin{array}{ccc}
\lb^8&\lb^6&\lb^4\\
\lb^6&\lb^4&\lb^2\\
\lb^4&\lb^2&1\\
\end{array}\right)
\ee
Therefore, this simple charge assignment  predicts also a
hierarchical texture for the left--handed neutrino Majorana mass.
The  mass scale is determined by the suppression mass-factor
 $\frac{m_W^2}{M_U}$ so there is a sufficient  suppression without the use
right-handed neutrino fields and  the see-saw mechanism. As in all
other matrices, only the third--generation diagonal coupling
$L_3L_3H_2^2$ appears at the tree-level.

This solution does not suppress $R$--parity violating couplings so one has to assume
that $R-$parity is a good symmetry. It does not also suppress the $\mu$--term so
one has to assume the existence of another mechanism that deals with this problem.

\item

{\bf Solution B}\\ Here we present another example which results to the same mass
matrices as in Solution A, and with similar suppression of dimension--5 operators. The
difference here is that
$L_iL_jH_2H_2$ operators are also suppressed (the stronger coupling is of the order
$\lb^{12}$). Similar comments with Solution A hold for the R--parity violating couplings and the $\mu$ term.
\end{itemize}

In the above two cases, we have used integer $U(1)_X$ charges for fermions and
higgs fields. Going further, we present few more examples where now we
introduce  fractional $U(1)_X$ charge assignments.

\begin{itemize}
\item
{\bf Solution C}\\ This solution gives mass matrices similar to
Solution A. The dimension--5 operators are suppressed according to
(\ref{sup5}). The difference here is that all R-parity violating
couplings in (\ref{blv},\ref{hmx}) vanish explicitly. However, the
Majoranna neutrino mass operator survives and takes the form
\be
L_i L_j H_2 H_2\sim \frac{m_W^2}{M_U} \left(\begin{array}{ccc}
\lb^3&\lb^1&\l^1\\
\lb^1&\l^1&\l^3\\
\l^1&\l^3&\l^5\\
\end{array}\right)
\ee
\item
{\bf Solution D}\\ The charges in this case appear also in Table
\ref{sols}.
 They yield up
quark fermion mass textures of the form
\be
m_U=\left(\begin{array}{ccc}
\eb^6&0&\eb^3\\
0&\eb^3&0\\ \eb^3&0&1\\
\end{array}\right)
\ee
and similarly for down quarks and leptons
\be
m_D=m_L=\left(\begin{array}{ccc}
\lb^6&0&\lb^3\\
0&\lb^3&0\\
\lb^3&0&1\\
\end{array}\right)
\ee
This solution completely eliminates all dimension--5 proton decay as well as all
$R$--parity violating couplings (\ref{blv}). $L_iL_jH_2H_2$ operators are also
suppressed. The only additional mechanism one needs is for the suppression of the
$\mu$ term as the solution belongs to the category $h_+=0$.

\item {\bf Solution E}\\
The quark and lepton matrices are
\be
m_U=\left(\begin{array}{ccc}
\eb^8&\eb^6&\eb^4\\
\eb^6&\eb^4&\eb^2\\
\eb^4&\eb^2&1\\
\end{array}\right),
\ee
and similarly for $m_D$ with $\eb\to\lb$, while the charged leptons are given by
\be
m_L = \left(\begin{array}{ccc}
\l^8&\l^2&\l^4\\
\l^2&\lb^4&\lb^2\\
\l^4&\lb^2&1\\
\end{array}\right)\,.
\ee
This Solution forbids all dimension--5 operators as
well as baryon and lepton number violating couplings (\ref{blv}). One concludes that
all additional dimension--5 operators are also suppressed whatever  the charges of
these operators are. At the same time it suppresses the higgs mixing as
 the $\mu$--term appears now through the
non--renormalizable term
\ba
W_{NR}\ra \left(\frac{\bar\phi}{M_U}\right)^{16}H_1 H_2,
\ea
therefore the higgs doublets are protected from receiving an unacceptable large mass.
Surprisingly the operator $L_i L_j H_2 H_2 $ is not suppressed; it gives a
left--handed Majorana neutrino texture,
\be
L_i L_j H_2 H_2\sim \frac{m_W^2}{M_U} \left(\begin{array}{ccc}
\l^9&\l^3&\l^5\\
\l^3&\lb^3&\lb^1\\
\l^5&\lb^1&\l^1\\
\end{array}\right)\,.
\ee
which exhibits the phenomenologically interesting feature of 
a rather large  mixing in the $\nu_\mu-\nu_\tau$-sector.
The price one has to pay for all these welcomed features are the rather
 exotic charges.
This is a feature also pointed out
 in \cite{Froggatt:1979nt,Binetruy:1995ru}.

\end{itemize}
\section{Conclusions}

In this work we have attempted to generate the hierarchical standard model fermion
mass spectrum by means of an anomalous abelian family symmetry $U(1)_X$ and in the
context of the minimal unification scenario.
We have extended previous analyses by considering the $U(1)_X$ to be
family dependent a possibility that naturally arises in superstring model building. A
minimum number of fields, --one singlet and its conjugate-- where used to break the
anomalous
$U(1)_X$ symmetry at a high scale. We have assumed that the $U(1)_X$ anomaly is
cancelled by the string Green--Schwarz anomaly cancellation mechanism. We have imposed
conditions on the
$U(1)_X$--matter and higgs charges by requiring symmetric mass matrices and
tree--level couplings for the third generation. We have demanded the mixed
$SU(3)^2U(1)_X$, $SU(2)^2U(1)_X$ and
$U(1)_Y^2U(1)_X$ anomalies to be proportional to the Kac--Moody constants
$k_3=k_2=3k_1/5=1$ as well as cancellation of the
$U(1)_Y^2 U(1)_X$ mixed and $U(1)_X^3$ anomalies. 
The general solution of the resulting equations
has been determined and all possible textures of the fermion mass matrices were
classified in terms of the admissible values of the sum of the two
$U(1)_X$--higgs charges. The cases of zero and integer values of the
higgs sum charge where considered while non-integer values, although possible, were
not discussed since they lead to a massless charged-lepton eigenstate and prevent the
appearance of a $\mu$--term to all orders. Using the freedom left by the anomaly
conditions on the
$U(1)_X$ charges, four distinct phenomenologically acceptable texture--zero
solutions for the fermion mass hierarchy problem have been predicted. The mass
hierarchy is determined from powers of parameters defined as the dimensionless
ratio of the singlet vevs over some high (string) scale. The magnitude of the
expansion parameters is constrained due to the $D$--term cancellation mechanism
which determines the singlet vevs in terms of the unification scale and the
common (string) coupling. We note that up and down quark mass matrices are
predicted to have the same form due to the initial assumptions that the matrices
are symmetric and the requirement that both top and bottom Yukawa couplings
appear at the tree--level. However, the predicted quark masses can be reconciled
with the low energy measured values due to the possible appearance of different
expansion parameters in the matrices and renormalization running effects. The
success of the above scenario might look more impressive if some of the
simplifying assumptions, were relaxed. Nevertheless, we find it remarkable that even
in this simple extension of the minimal supersymmetric standard model one may predict
to a good approximation the big mass gaps observed in the particle spectrum.
It is tempting to extend the analysis by relaxing some of the unnecessary assumptions
and re--examine the above model.

The rather remarkable fact is that this simple $U(1)_X$ anomalous symmetry with the
constraints implied by the anomaly cancellation conditions, allow fermion charge
assignments which can suppress, --or in certain cases-- eliminate all dangerous baryon
and lepton number violating operators. In the case that $R$--parity is a good symmetry
we have found solutions that can suppress the dangerous dimension--five proton decay
operators (allowed by $R$--parity). We have also found solutions that do not need the
introduction of $R$--parity since there,  all R--parity violating
couplings are naturally suppressed.

We have further shown that these solutions may also suppress sufficiently the higgs
doublets mixing parameter ($\mu$--term) and keep them massless down to the electroweak
scale. This latter possibility requires the introduction of a rather big charge for the
sum of the higgs doublets which demand rather peculiar $U(1)_X$ assignments for MSSM
fields at least in the case that the singlet charge is unity. In the context of these
solutions we have also succeeded to find cases which
 provide the left--handed neutrino with
acceptable   Majorana masses.

It is remarkable that most or all of the good features mentioned above can occur
simultaneously in a few simple solutions which we presented in this work. 

\vspace*{2cm}
\noindent
{\bf Acknowledgements} ~~\\
\noindent
The work of J.R. was supported in part by the EU under the TMR
contract ERBFMRX-CT96-0090.



\newpage

\begin{thebibliography}{99}

\bibitem{Ramond:1993kv}
P.~Ramond, R.G.~Roberts and G.G.~Ross, Nucl. Phys. {\bf B406} (1993) 19
hep-ph/9303320.

\bibitem{Green:1984sg}
M.B.~Green and J.H.~Schwarz, Phys. Lett. {\bf 149B} (1984) 117.


\bibitem{Froggatt:1979nt}
C.D.~Froggatt and H.B.~Nielsen, Nucl. Phys. {\bf B147} (1979) 277.

\bibitem{Ibanez:1993fy}
L.E.~Ibanez, Phys. Lett. {\bf B303} (1993) 55 hep-ph/9205234.

\bibitem{Ibanez:1994ig}
L.~Ibanez and G.G.~Ross, Phys. Lett. {\bf B332} (1994) 100 hep-ph/9403338.

\bibitem{Leurer:1994gy}
M.~Leurer, Y.~Nir and N.~Seiberg, Nucl. Phys. {\bf B420} (1994) 468 hep-ph/9310320.

\bibitem{Binetruy:1995ru}
P.~Binetruy and P.~Ramond, Phys. Lett. {\bf B350} (1995) 49 hep-ph/9412385;\\
P.~Binetruy, S.~Lavignac and P.~Ramond,
Nucl.\ Phys.\ {\bf B477} (1996) 353 hep-ph/9601243.


\bibitem{Nir}
Y. Nir, Phys. Lett. {\bf B354} (1995) 107;\\
 Y.~Grossman and Y.~Nir,
Nucl.\ Phys.\ {\bf B448} (1995) 30 hep-ph/9502418.
\bibitem{Dreiner:1995ra}
H.~Dreiner, G.K.~Leontaris, S.~Lola, G.G.~Ross and C.~Scheich,
Nucl.\ Phys.\ {\bf B436} (1995) 461
hep-ph/9409369.

\bibitem{Dudas:1995yu}
E.~Dudas, S.~Pokorski and C.A.~Savoy,
Phys.\ Lett.\ {\bf B356} (1995) 45
hep-ph/9504292.

\bibitem{Altarelli:1998ns}
G.~Altarelli and F.~Feruglio,
Phys.\ Lett.\ {\bf B451} (1999) 388
hep-ph/9812475.

\bibitem{DSW}
M.~Dine, N.~Seiberg and E.~Witten, Nucl. Phys. {\bf B289} (1987) 589.

\bibitem{FI}
P.~Fayet and J.~Iliopoulos, Phys. Lett. {\bf B51} (1974) 461.



\bibitem{ADS}
J.J.~Atick, L.J.~Dixon and A.~Sen, Nucl. Phys. {\bf B292} (1987) 109.


\bibitem{Leontaris:1993wp}
G.K.~Leontaris and J.D.~Vergados, Phys. Lett. {\bf B305} (1993) 242 hep-ph/9301291.


\bibitem{Ellis:1997ni}
J.~Ellis, G.K.~Leontaris, S.~Lola and D.V.~Nanopoulos, Phys. Lett.
{\bf B425} (1998) 86 hep-ph/9711476.


\bibitem{Giudice:1992an}
G.F.~Giudice, Mod. Phys. Lett. {\bf A7} (1992) 2429
hep-ph/9204215.

\bibitem{Georgi:1979df}
H.~Georgi and C.~Jarlskog, Phys. Lett. {\bf 86B} (1979) 297.



\bibitem{Farrar:1978xj}
G.R.~Farrar and P.~Fayet, Phys. Lett. {\bf 76B} (1978) 575.


\bibitem{Farrar:1983te}
G.R.~Farrar and S.~Weinberg, Phys. Rev. {\bf D27} (1983) 2732.

\bibitem{IR}
L.E.~Ibanez and G.G.~Ross, Nucl.\ Phys.\ {\bf B368} (1992) 3.


\bibitem{Ellis:1999ce}
J.~Ellis, G.K.~Leontaris and J.~Rizos, hep-ph/9907476,
to appear in  Phys.\ Lett. \ {\bf B}.

\bibitem{Barbieri:1980hc}
R.~Barbieri, J.~Ellis and M.K.~Gaillard,
Phys.\ Lett.\ {\bf 90B} (1980) 249.


\bibitem{PQ}
R.D.~Peccei and H.R.~Quinn,
Phys.\ Rev.\ Lett.\ {\bf 38} (1977) 1440.

\bibitem{Ibanez:1999it}
L.E.~Ibanez and F.~Quevedo, hep-ph/9908305.

\bibitem{Ellis:1998rj}
J.~Ellis, S.~Lola and G.G.~Ross,
Nucl.\ Phys.\ {\bf B526} (1998) 115 hep-ph/9803308.






\end{thebibliography}
\end{document}